\documentclass[preprint]{aastex63}

\usepackage{graphicx}
\usepackage{amsmath}
\usepackage[caption=false]{subfig}

\begin{document}


\title{The Solar-wind with Hydrogen Ion Exchange and Large-scale Dynamics (SHIELD) model: A Self-Consistent Kinetic-MHD Model of the Outer Heliosphere}


\author{A. T.  Michael}
\affil{Astronomy Department, Boston University,
    Boston, MA 02115}
\email{atmich@bu.edu}    

\author{M. Opher}
\affil{Astronomy Department, Boston University,
    Boston, MA 02115}

\author{G. T\'{o}th}
\affil{University of Michigan, Ann Arbor, MI 48109}

\author{V. Tenishev}
\affil{University of Michigan, Ann Arbor, MI 48109}


\author{D. Borovikov}
\affil{University of Michigan, Ann Arbor, MI 48109}



\begin{abstract}
Neutral hydrogen has been shown to greatly impact the plasma flow in the heliopshere and the location of the heliospheric boundaries. We present the results of the Solar-wind with Hydrogen Ion Exchange and Large-scale Dynamics (SHIELD) model, a new, self-consistent, kinetic-MHD model of the outer heliosphere within the Space Weather Modeling Framework. The charge-exchange mean free path is on order of the size of the heliosphere; therefore, the neutral atoms cannot be described as a fluid. The SHIELD model couples the MHD solution for a single plasma fluid to the kinetic solution from for neutral hydrogen atoms streaming through the system. The kinetic code is based on the Adaptive Mesh Particle Simulator (AMPS), a Monte Carlo method for solving the Boltzmann equation. The SHIELD model accurately predicts the increased filtration of interstellar neutrals into the heliosphere. In order to verify the correct implementation within the model, we compare the results of the SHIELD model to other, well-established kinetic-MHD models. The SHIELD model matches the neutral hydrogen solution of these studies as well as the shift in all heliospheric boundaries closer to the Sun in comparison the the multi-fluid treatment of the neutral hydrogen atoms. Overall the SHIELD model shows excellent agreement to these models and is a significant improvement to the fluid treatment of interstellar hydrogen.
\end{abstract}


\section{Introduction}
Interstellar neutral hydrogen plays a major role in our understanding of the heliosphere. The heliosphere forms as the solar wind expands into the ionized portion of the local interstellar medium (ISM). The neutral atoms do not feel the boundaries separating the solar and interstellar plasma, instead stream into the the inner Solar System. Hydrogen atoms are the dominant component of the ISM and are coupled to the plasma through resonant charge exchange \citep{wallis1975}. This removes up to 30\% of momentum from the solar wind, decelerating the flow as it expands into the ISM. The influence of interstellar neutral hydrogen reduces the distance to the heliospheric boundaries by $\sim 150$ astronomical units (AU) and significantly alters the structure of the termination shock (TS), the location where the solar wind becomes subsonic \citep{baranov1993}.

Neutral hydrogen also provides a mechanism to observe the global structure of the heliosphere. As a result of charge exchange, energetic neutral atoms (ENAs) are produced with the velocity of their parent ions. ENAs travel relatively unimpeded through the heliosphere and can be detected as far in as 1 AU. ENA observations, such as those done by the Interstellar Boundary Explorerer (IBEX) \citep{schwadron2011,schwadron2014} and the Ion and Neutral Camera (INCA) instrument on \textit{Cassini} \citep{krimigis2009}, are able to indirectly probe the global structure of the heliosphere. Since the neutrals play such a significant role in the system, including them self-consistently into models is crucial to understand the global ENA maps and the in situ measurements of the plasma at \textit{Voyager} 1 and 2. 

The solar wind - ISM interaction can be separated into 4 regions. The heliosphere consists of the supersonic solar wind in the inner heliosphere and the subsonic solar wind in the heliosheath (HS). If the Sun's motion through the ISM is supersonic, the pristine ISM will become subsonic at the bow shock (BS), before being diverted at the heliopause (HP). \citet{baranov1981} was the first work to incorporated neutral hydrogen into computational models of the heliosphere. This model, as well as \citet{pauls1995}, treated the plasma and neutral hydrogen as two separate fluids, coupled by source terms which approximate the effect of charge exchange. Charge exchange in each region of the heliosphere will produce ENAs with distinct characteristics and charge exchange mean free paths. Each region has different characteristic parameters, therefore, a single fluid description for the neutrals is inadequate. 

Subsequent models have employed a multi-fluid approximation for the neutrals, defining a fluid population for each region of the heliosphere \citep{zank1996,williams1997,fahr2000,pogorelov2005,opher2009}. The multi-fluid approach for the neutrals is much more computationally efficient \citep{mcnutt1998}. Furthermore, \citet{alexashov2005} was the first to show that the plasma solution differs only by a few percent in the upwind direction in comparison to a kinetic treatment of the neutrals. Subsequent models that used a kinetic solution for the neutrals confirmed this result as well \citep{heerikhuisen2006,muller2008,alouani2011}. While it provides an incorrect solution for the neutral component, the lower computational cost of the multi-fluid approximation makes it advantageous when studying dynamical plasma processes such as the transient shock propagation \citep{provornikova2013}, solar cycle effects \citep{pogorelov2013,michael2015}, or turbulence. The fluid description assumes that there are sufficient H-H collisions to thermalize their distribution into a Maxwellian. However, H-H collisions are negligible throughout the heliosphere \citep{izmodenov2000}, causing the distribution functions to be distinctly non-Maxwellian \citep{izmodenov2001}. Additionally, since the charge exchange mean free path is on the order of the size of the heliosphere, the Knudson number, $Kn=l_{mfp}/L$ where $l_{mfp}$ is the mean free path and $L$ is the characteristic length scale of the system, for neutral hydrogen is around one. Therefore, a kinetic treatment is required to accurately model the neutrals \citep{izmodenov2000}.

The Monte Carlo method is one of the standard methods for numerically solving multidimensional kinetic equations. \citet{bird1994} has adapted it to model a rarefied gas whose dynamics are determined through a finite number of collisional processes such that the flow is unsteady. A main advantage of the Monte Carlo method is that it does not require the use of integro-differential equations that describe the evolution of the distribution function. Instead, the Monte Carlo method follows the trajectories of many simulated macro-particles throughout the domain. Collisions are handled through particle pairs rather than integrating the Boltzmann collision integral. This allows for a wider range of collisional processes to be included and makes it a natural choice for the outer heliosphere application.  

The first self-consistent kinetic treatment of the heliospheric neutrals was introduced by \citet{malama1991} and implemented by \citet{baranov1993}. In their model, \citet{baranov1993} solve the Boltzmann equation for the trajectory of the neutrals using a Monte Carlo approach. The Monte Carlo model was coupled to a single hydrodynamic plasma fluid which was also done in later models by \citet{muller2000} and \citet{heerikhuisen2006} as well. Charge exchange collisions are handled on a particle by particle basis while the source terms of the local plasma are the accumulation of events that occur in a particular location. The single plasma fluid was extended to also include helium ions and alpha particles \citep{izmodenov2003}  while anomalous cosmic rays have been treated as a massless, diffusive fluid by \cite{alexashov2004}. The Monte Carlo code can also be used to model additional interstellar atoms such as oxygen and nitrogen \citep{izmodenov2004}, as well as interstellar dust \citep{alexashov2016}. 

The solar magnetic field has been shown to significantly impact the location of the heliospheric boundaries \citep{izmodenov2015}. \citet{opher2015} also show that the solar magnetic field can collimate the solar wind in the heliosheath (HS), possibly resulting in a two-lobe structure of the heliotail. An analytic solution for the collimation of the solar wind by the magnetic field was developed by \citet{drake2015}. They showed that the solar wind magnetic field strength affect the strength of the lobes as well as the thickness of the HS. Therefore, the heliosphere cannot be modeled hydrodynamically. A self-consistent model of the heliosphere with kinetic neutrals coupled to a single magnetohydrodynamic (MHD) plasma fluid has been developed by \citet{pogorelov2009} and \citet{izmodenov2015}. These kinetic-MHD models have been very successful at modeling the outer heliosphere, however, they do have some limitations. \citet{opher2017} showed that magnetic reconnection can occur in the eastern flank of the heliosphere which organizes the magnetic field ahead of the HP. \citet{izmodenov2015} use a scheme that restricts any communication between the ISM and solar wind across the HP to limit non-ideal MHD effects and therefore cannot model this behavior. Additionally, \citet{pogorelov2009} use a dipole description of the solar magnetic field which induces large numerical dissipation of the magnetic field and numerical reconnection across the HP. Since the scale of the current sheet is much smaller than the heliosphere, it is extremely difficult to resolve the structure. This results in the removal of magnetic field strength over large regions of HS, making the results strongly affected by numerical artifacts \citep{michael2018}.

In this work, we introduce the Solar-wind with Hydrogen Ion Exchange and Large-scale Dynamics (SHIELD) model. A new global, self-consistent kinetic-MHD model of the heliosphere that utilizes the MHD solution of \citet{opher2015}. The SHIELD model couples the Outer Heliosphere (OH) and Particle Tracker (PT) components within the Space Weather Modeling Framework (SWMF) \citet{toth2005}. The individual models are described in section 2. The numerical scheme of the Monte Carlo code used in the PT component is presented in section 2.1. Section 2.2 details how the Monte Carlo model is applied it to the outer heliosphere and coupled to the MHD solution. The verification of our model is presented in section 2.4. Here, we adopt the same boundary conditions used in both \citet{alexashov2005} and \citet{heerikhuisen2006} to compare our solution directly with their results. Finally, the conclusions are presented in section 4. 

\section{Model}
Our model couples the OH and PT components within the SWMF. The SWMF is a framework developed for physics-based space weather modeling. Frameworks are advantageous tools for space weather due to the vast spatial and temporal scales needed to model the physical processes associated with the Sun-Earth system. The SWMF allows for each physical domain to be described by separate models which can apply the correct discretization needed for that region. Each physical domain is contained within a different component which are coupled to each other through the framework. The SWMF compiles the desired models into a single executable and distributes the components onto a parallel computer, where they are executed simultaneously and coupled efficiently throughout the run. Therefore, the SWMF is capable of self-consistently modeling physical processes including the ejection of a coronal mass ejection from the solar surface to its impact on the magnetosphere and generated currents in the ionosphere.   

The OH component is based on the Block-Adaptive Tree Solar wind Roe-Type Upwind Scheme (BATS-R-US) solver, a highly parallel, 3-dimensional, block-adaptive, upwind finite-volume MHD code \citep{toth2012}. BATS-R-US was first adapted for the outer heliosphere by \citet{opher2003}. \citet{opher2003} instituted a 3D MHD solution that modeled the plasma component as a single fluid and neglected the interstellar magnetic field and neutrals. The OH model was further developed by \citet{opher2004} to include both the solar and interstellar magnetic field, making the model a full 3D ideal MHD simulation. \citet{opher2009} was the first to incorporate the interstellar neutral hydrogen atoms into the model with the four fluid approximation. When run in stand-alone mode, the OH component is a global 3D multi-fluid MHD simulation of the outer heliosphere that describes the plasma and four neutral hydrogen species. The OH stand-alone model solves the ideal MHD equations for the plasma and a separate set of Euler's equations for the different populations of neutral atoms. The neutral fluids are coupled to the plasma through source terms resulting from charge exchange as calculated by \citet{mcnutt1998}. Subsequent additions to the \citet{opher2009} model includes the sector structure of the heliospheric current sheet \citep{opher2011} along with time and latitude-dependent solar cycle variations of the solar wind \citep{provornikova2014,michael2015}. 

The OH component has been updated to include the pick-up ions (PUIs) as a separate fluid \citep{opher2019}. \citet{opher2019} solves the coupled, multi-fluid MHD equations for the thermal solar wind ions and PUIs as developed by \citet{glocer2009} and \citet{toth2012}. The plasma components and neutral fluids are coupled through updated source terms depending on which fluids are undergoing charge exchange \citep{opher2019}. In this paper, we chose to couple the kinetic model to the single plasma fluid approximation. Future extensions of SHIELD will couple the kinetic code to the multi-ion model of \citet{opher2019}. This is in accordance with other kinetic-MHD models of the outer heliosphere of \citet{baranov1993} and \citet{heerikhuisen2006} and allows for a model verification before advancing to the the multi-ion coupling.

The PT component is the Adaptive Mesh Particle Simulator (AMPS) code, a global, kinetic, 3D kinetic particle code developed within the framework of the Direct Simulation Monte Carlo methods (DSMC) for the purpose of solving the Boltzmann equation for the motion and interaction of a dusty, partially ionized, multi-species gas in cometary comae \citep{tenishev2008}. Among others, AMPS is capable of modeling collisions between species, photodissociation, energy exchange between internal vibrational and rotational states, and radiative cooling. AMPS employs a block-adaptive mesh that can utilize an AMR procedure to ensure that the cells are refined to resolve the local mean free path and capture the geometry of complex systems. The PT component has been applied to cometary comae \citep{tenishev2008}, the Moon's exosphere \citep{tenishev2014}, the Martian ionosphere and upper atmosphere \citep{valeille2009,lee2014,dong2018}, as well as electron acceleration during solar flares \citep{borovikov2017}. A Monte Carlo method was also used by \citet{heerikhuisen2006} and we use a similar method here to solve for the motion and distribution of interstellar neutral hydrogen.

In the SHIELD model, we couple the OH and PT components. The PT component is applied to the outer heliosphere and is used to kinetically solve for the neutral atom trajectories, including charge-exchange collisions with the plasma. The SHIELD model treats the plasma as a single fluid and couples the MHD solution with the kinetic solution of the neutrals to form a self-consistent model of the heliosphere. In the subsequent sections, we detail the numerical scheme used by AMPS as well as the methods for how it is applied to model the outer heliosphere. 

\subsection{Numerical Scheme of AMPS}
AMPS is a 3D direct simulation Monte Carlo (DSMC) model developed to solve the Boltzmann equation for a dusty, partially ionized, multi-species cometary comae. DSMC models are very effective at modeling collisional neutral fluids kinetically, which makes them a natural choice to model the neutrals streaming throughout the heliosphere. 

The DSMC method employed by AMPS uses a series of Markov chains to solve for the evolution of the system \citep{tenishev2008}. AMPS accomplishes this by describing the neutral flow with a large but finite set of model particles. The number of interstellar neutrals traveling throughout the heliosphere is very large, therefore, a model particle within AMPS represents many real particles. The basic idea of the employed method is the separation of the translational motion of the neutrals through the heliosphere from their interaction with the plasma of the ambient solar wind. The latter is simulated probabalistically, calculating the time interval between successive charge exchange events in a particular computational cell according to the Poisson distribution. For modeling convection in the simulated environment, model particles are affected by the same physical laws as real neutrals in the heliosphere. That is equivalent to solving the kinetic equation with the method of characteristics. The model particle will follow the same trajectory as the real atoms would. 

The velocity of the model particles injected into the computational domain are distributed according to the distribution function of real interstellar neutral atoms ahead of the heliosphere. AMPS simulates transport of these particles through the domain, tracking their trajectories through the heliosphere and determines the probability that a collision occurs for each model particle. If a collision does occur for a particle, the velocity is updated appropriately as well as the particle weight if there is a loss of real atoms during the event. After the collisions, forces are applied to the particles and their locations are updated accordingly. The particles can be sampled at any time, producing the distribution function at any location. 

\subsection{Application to the Outer Heliosphere}
AMPS solves the time-dependent Boltzmann equation
\begin{equation}
    \frac{\partial f_s}{\partial t}+\mathbf{v_s} \cdot \nabla f_s + \frac{\mathbf{F}}{m_s} \cdot \nabla_{\mathbf{v_s}} f_s = \left (\frac{\partial f_s}{\partial t} \right )_{collision} 
    \label{eq:Boltzmann}
\end{equation}
for species s, with a distribution function, $f_s$. Here \textbf{F} is an external macroscopic force and $(\partial f_s/\partial t )_c $ is the effect on the distribution function due to collisions as well as stochastic interactions with the ambient media. 

The neutral composition of the ISM can be determined, as neutrals are unaffected by the heliospheric boundaries and penetrate deep into the heliosphere. These atoms, such as helium atoms that interact very weakly through charge exchange, can be measured directly by instruments such as IBEX \citep{mobius2012}, or indirectly, as they are eventually ionized and picked up by the solar wind \citep{rucinski1996}. Through these methods, H, He, N, O, Ne, and Ar have been detected. Neutral hydrogen constitutes 92\% of the ISM by number. The rest of the ISM is comprised mostly of He atoms with N, O, Ne, and Ar comprising fractions of a percent \citep{gloeckler2001}. Since helium atoms very rarely undergo charge exchange with ionized hydrogen, their impact on the structure of the heliosphere is minimal. Solar wind protons readily charge exchange with oxygen atoms, however, the charge exchange cross section is an order of magnitude smaller than the H-H$^+$ process for energies typical of the solar wind \citep{lindsay2005}. Additionally, the density of oxygen is many orders of magnitude below the H density in the ISM, causing this reaction be much more infrequent. We therefore model the neutral component of the ISM as a single hydrogen species gas and neglect all other species.

\citet{izmodenov2000} have shown that elastic hydrogen-hydrogen or hydrogen-proton collisions are negligible over the spatial scale of the heliosphere when compared to the charge exchange mean free path. Photoionization is an additional loss source for the neutrals, however, the solar wind boundary conditions are incorporated at 30 au and extrapolated into the origin using scaling relations. At 30 au, the charge exchange rate is several orders of magnitude larger than photoionization, therefore, the photoionization process is neglected within our calculation. Hot electrons in the HS could also contribute to ionization of neutral atoms \citep{gruntman2015}, however, Voyager 2 observed that electrons have energies below the 10 eV instrument threshold \citep{richardson2008}. These electrons are not hot enough for electron impact ionization to have an influence on the mass loading of the solar wind in the HS, thus we have not included this process within SHIELD. As a result, the only collisional process we model is charge exchange. The resulting collisional operator can be described in terms of production and loss of neutral atoms
\begin{multline}
       \left (\frac{\partial f_s}{\partial t} \right )_{collision} = f_p(\mathbf{x},\mathbf{v},t)\int |\mathbf{v}_H-\mathbf{v}|~ \sigma_{ex}(|\mathbf{v}_H-\mathbf{v}|)~f_H(\mathbf{x},\mathbf{v}_H,t) ~d\mathbf{v}_H \\
   - f_H(\mathbf{x},\mathbf{v},t)\int |\mathbf{v}-\mathbf{v}_p| ~\sigma_{ex}(|\mathbf{v}-\mathbf{v}_p|) ~f_p(\mathbf{x},\mathbf{v}_p,t) ~d\mathbf{v}_p
   \label{eq:collision}
\end{multline}
where $f_p(\mathbf{x},\mathbf{v},t)$ and $f_H(\mathbf{x},\mathbf{v},t)$ are the distribution functions of the plasma and the neutrals at a specific location, and $\sigma_{ex}$ is the charge exchange cross section, calculated using the \citet{lindsay2005} approximation,
\begin{equation}
    \sigma_{ex}(\Delta E) = (4.15-0.531 ln(\Delta E))^2 (1-e^{-67.3/\Delta E})^{4.5} cm^2
    \label{eq:LSsigma}
\end{equation}
Here $\Delta E$ is the relative proton-neutral atom energy in keV. 

The external forces acting on the neutrals streaming into the heliosphere are the solar radiation pressure and gravity. At distances larger than where our inner boundary is applied, these forces are negligible. Consequently, there is no net external force acting on the atoms, allowing the last term on the left-hand side of equation \ref{eq:Boltzmann} to be dropped. This significantly reduces the complexity of the Boltzmann equation. The atoms travel on straight-line trajectories until a charge exchange event occurs and the velocity of the particles are altered. 

\begin{figure}[!ht]
\centering
\subfloat[]{%
  \includegraphics[width=0.4\textwidth]{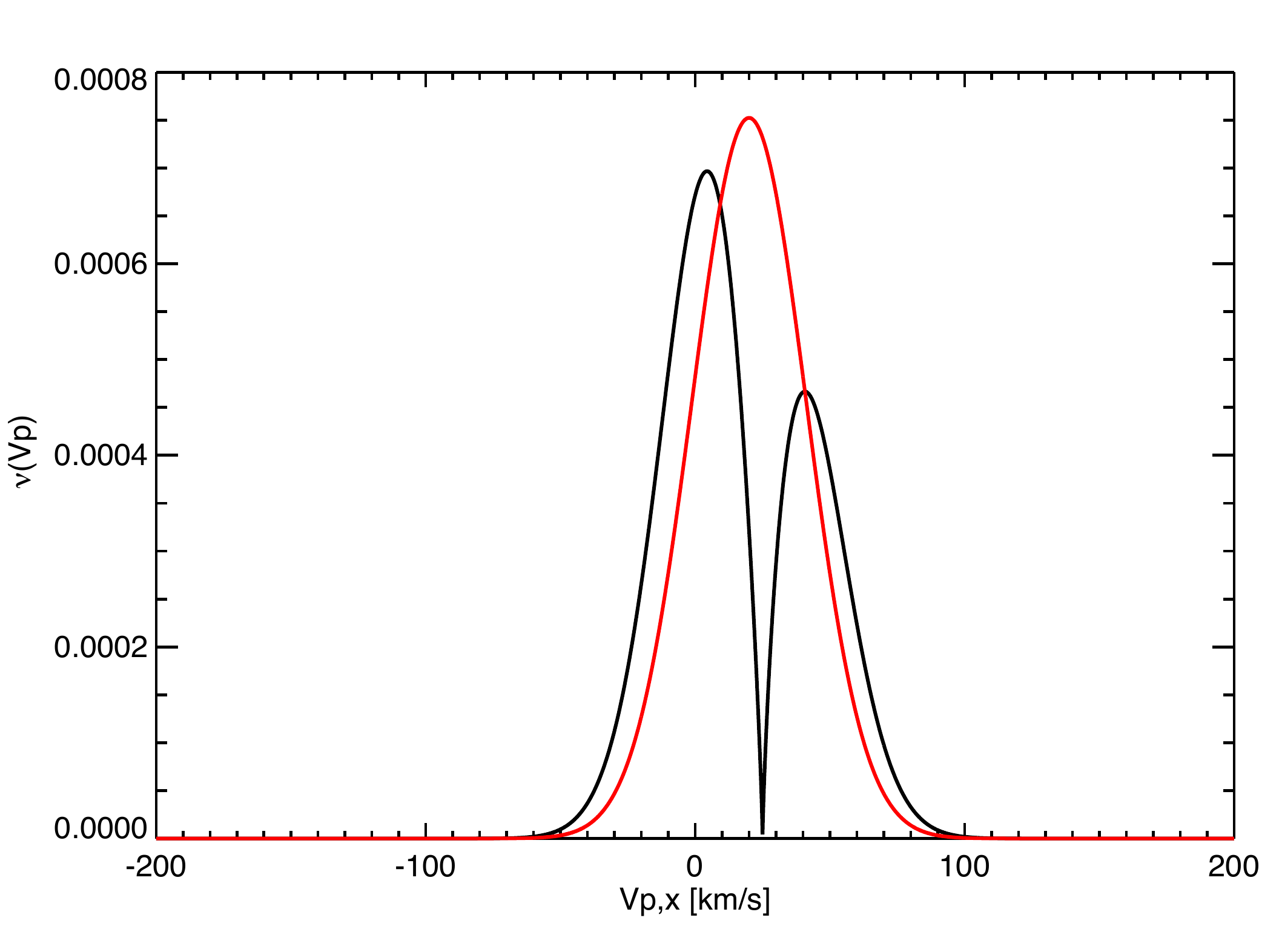}
  \label{fig:1dvpselect}%
}
\subfloat[]{%
  \includegraphics[width=0.55\textwidth]{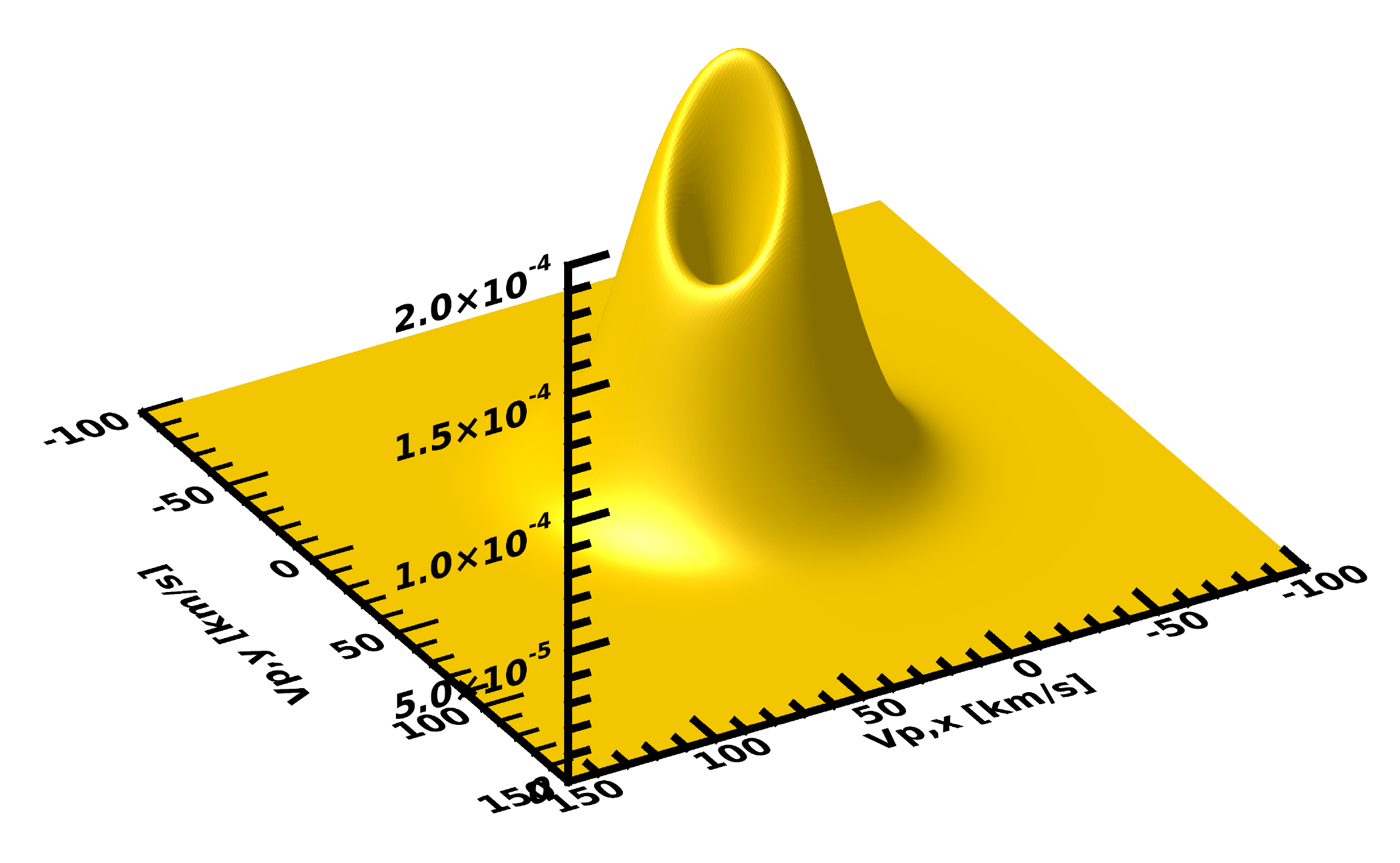}
  \label{fig:2dvpselect}%
}
\caption{1D \protect\subref{fig:1dvpselect} and 2D \protect\subref{fig:2dvpselect} frequency distribution functions of the selected proton from the local Maxwellian plasma distribution as well as the resulting ENA during a charge exchange event. Here the initial neutral particle has $\mathbf{v}_H = (25,25)$ km/s and the bulk plasma velocity and thermal velocity are $\mathbf{v}_p = (20,20)$ km/s and $v_{th,p}=30$ km/s, respectively. The Maxwellian distribution is shown in red in \protect\subref{fig:1dvpselect} for comparison. \label{fig:vpselect}}
\end{figure}

During a charge exchange event, a proton must be selected from the local plasma Maxwellian distribution. Neutral atoms more frequently undergo charge exchange when the velocity difference between the proton and the neutral particle is maximized while charge exchange with solar wind particles whose velocity is near $\mathbf{v}_H$ rarely occurs. That causes a depletion in the distribution around the neutral atom velocity, $\mathbf{v}_H$. The frequency distribution function of a proton with velocity, $\mathbf{v}_p$, is given by
\begin{equation}
    \nu(\mathbf{v}_p) \sim |\mathbf{v}_H-\mathbf{v}_p|~\sigma_{ex}(|\mathbf{v}_H-\mathbf{v}_p|)~ e^{-(\mathbf{v}_p-\mathbf{u}_p)^2/v_{th,p}^2}
    \label{eq:vpfreqdist}
\end{equation}
where $\mathbf{v}_H$ is the initial hydrogen atom undergoing charge-exchange, $\mathbf{u}_p$ is the bulk plasma velocity, and $v_{th,p}=\sqrt{2kT_p/m_H}$ is the thermal speed of the plasma, as is done in \citet{malama1991} and a similar approach is used in \citet{heerikhuisen2006}. An example 2D frequency distribution function, a 1D cut through the center of the distribution, as well as a comparison to the original Maxwellian distribution of the plasma are presented in Figure \ref{fig:vpselect}. In this example, the local bulk plasma velocity, $\mathbf{u_p}$, is less than the neutral hydrogen atom speed. Solar wind protons with velocities less than the neutral hydrogen atom's initial velocity are therefore more likely to undergo charge exchange than those moving faster than the neutral atom. The opposite is also true. If the local bulk plasma velocity is larger than the neutral hydrogen atom velocity, the probability to undergo charge exchange is larger for protons with velocities larger than $\mathbf{v_H}$. 

Once the velocity of a solar wind proton participating in the charge exchange event is determined, the resulting ENA is generated and assigned the selected proton velocity. The region of the heliosphere which the ENA was generated in is also retained to track the propagation of each population of neutrals. The atoms can be sampled for each region to produce a solution for each population to determine the dominant population in each location as well as an easy comparison to the multi-fluid approach. 1D and phase space distribution functions can also be produced for the total solution or for each population to determine how the populations evolve throughout the domain.  

The impact of the charge exchange process on the plasma is approximated through source terms to the continuity, momentum, and energy equations of the ideal MHD system.
\begin{gather}
   \frac{\partial \rho}{\partial t} + \nabla \cdot (\rho \mathbf{u}) = S_{\rho} \\
   \frac{\partial (\rho \mathbf{u})}{\partial t} + \nabla \cdot \left [\rho \mathbf{u}\mathbf{u} + \left (p + \frac{B^2}{2\mu_0} \right ) \cdot \mathbf{I} - \frac{\mathbf{B}\mathbf{B}}{4\pi} \right ] = \mathbf{S}_{\rho v} \\
   \frac{\partial \epsilon}{\partial t} + \nabla \cdot \left [\mathbf{u} \left (\epsilon + p + \frac{B^2}{2\mu_0} \right ) - \frac{(\mathbf{u}\cdot\mathbf{B})\mathbf{B}}{\mu_0} \right ] = S_{\epsilon}, \\
   \epsilon = \frac{\rho u^2}{2} + \frac{p}{\gamma-1} +\frac{B^2}{2\mu_0}
\end{gather}
In the single fluid approximation, where the newly formed pick up ion is immediately assimilated into the plasma, charge exchange does not alter the plasma's density. However, since the pick up ion has a different initial velocity than the original proton, the collision alters the momentum and energy of the plasma as the picked up ion is instantaneously accelerated to the plasma velocity. The change in momentum and energy due to charge exchange can be found by evaluating the moments of the collisional term (Eq. \ref{eq:collision})
\begin{equation}
\begin{aligned}
   S_\rho & = 0 \\ 
   \mathbf{S}_{\rho v} & = \int \int m_p~|\mathbf{v}_H-\mathbf{v}_p|~\sigma_{ex}(|\mathbf{v}_H-\mathbf{v}_p|)~(\mathbf{v}_H-\mathbf{v}_p)~f_H(\mathbf{x},\mathbf{v}_H,t)~f_p(\mathbf{x},\mathbf{v}_p,t) ~d\mathbf{v}_H ~d\mathbf{v}_p \\
   S_\epsilon & = \int \int m_p~|\mathbf{v}_H-\mathbf{v}_p|~\sigma_{ex}(|\mathbf{v}_H-\mathbf{v}_p|)~\left(\frac{\mathbf{v}^2_H}{2}-\frac{\mathbf{v}^2_p}{2}\right)~f_H(\mathbf{x},\mathbf{v}_H,t)~f_p(\mathbf{x},\mathbf{v}_p,t) ~d\mathbf{v}_H ~d\mathbf{v}_p 
\end{aligned}
\end{equation}

By assuming the distribution function of the neutral atoms is also a Maxwellian distribution, analytic approximations can be found for \textbf{S}$_{\rho v}$ and S$_\epsilon$ \citep{pauls1995,mcnutt1998,williams1997}. This allows for a fast and efficient calculation of the source terms which are updated every time step and results in a quicker convergence of the solution for both steady-state and time-dependent problems. The DSMC model, however, allows for any form of the neutral distribution function. Following the work of \citet{malama1991}, a statistical estimation of the source terms can be found by summing the changes in momentum and energy from individual charge exchange events in each cell over a given time interval. With this approach, the source terms within a cell `k' can be expressed as
\begin{equation}
\begin{aligned}
   S_\rho & = 0 \\ 
   \mathbf{S}_{\rho v} & = \frac{1}{V_k~\Delta t} \displaystyle\sum_{i=1}^{n_{ex}}~\mu_i m_i ~(\mathbf{v}_{H,i}- \mathbf{v}_{p,i}) \\
   S_\epsilon & =  \frac{1}{V_k~\Delta t} \displaystyle\sum_{i=1}^{n_{ex}}~\frac{\mu_i m_i}{2} ~(\mathbf{v}^2_{H,i}- \mathbf{v}^2_{p,i})
\end{aligned}
\end{equation}
where $\mu_i$ is the particle weight of the atom, $V_k$ is the volume of the cell, $\Delta t$ is the time interval over which the charge exchange events occurred, n$_{ex}$ is the number of charge exchange events that occurred in cell over that time span, $\mathbf{v}_{H,i}$ is the velocity of the original hydrogen atom, and $\mathbf{v}_{p,i}$ is the velocity of the initial proton selected according to the frequency distribution function in Eq. \ref{eq:vpfreqdist}. In the example presented in Figure \ref{fig:1dvpselect}, the plasma bulk velocity is less than the neutral atom velocity. Protons with velocity near 0 km/s are much more probable to charge exchange with this neutral. This results in an increase in energy for the plasma and that is characteristic in the disturbed ISM region of the outer heliosphere. In the supersonic solar wind, the bulk plasma velocity is larger than the majority of parent neutral atom velocities. When this occurs, the most probable partner for the neutral to have a larger comparable speed, leading to a removal of momentum and energy from the plasma.   

The source terms need to be smooth from cell to cell in order for the resulting MHD solution to be stable. In order for this to occur, there needs to be enough charge exchange events, n$_{ex}$, within each cell to acquire accurate statistics. This can be achieved by increasing either the length of the time interval, $\Delta t$, before sampling or by increasing the total number of particles within the domain.

\subsection{Solving the Self-Consistent Problem}
\begin{figure}[!ht]
\begin{center}
\includegraphics[width=0.98\textwidth]{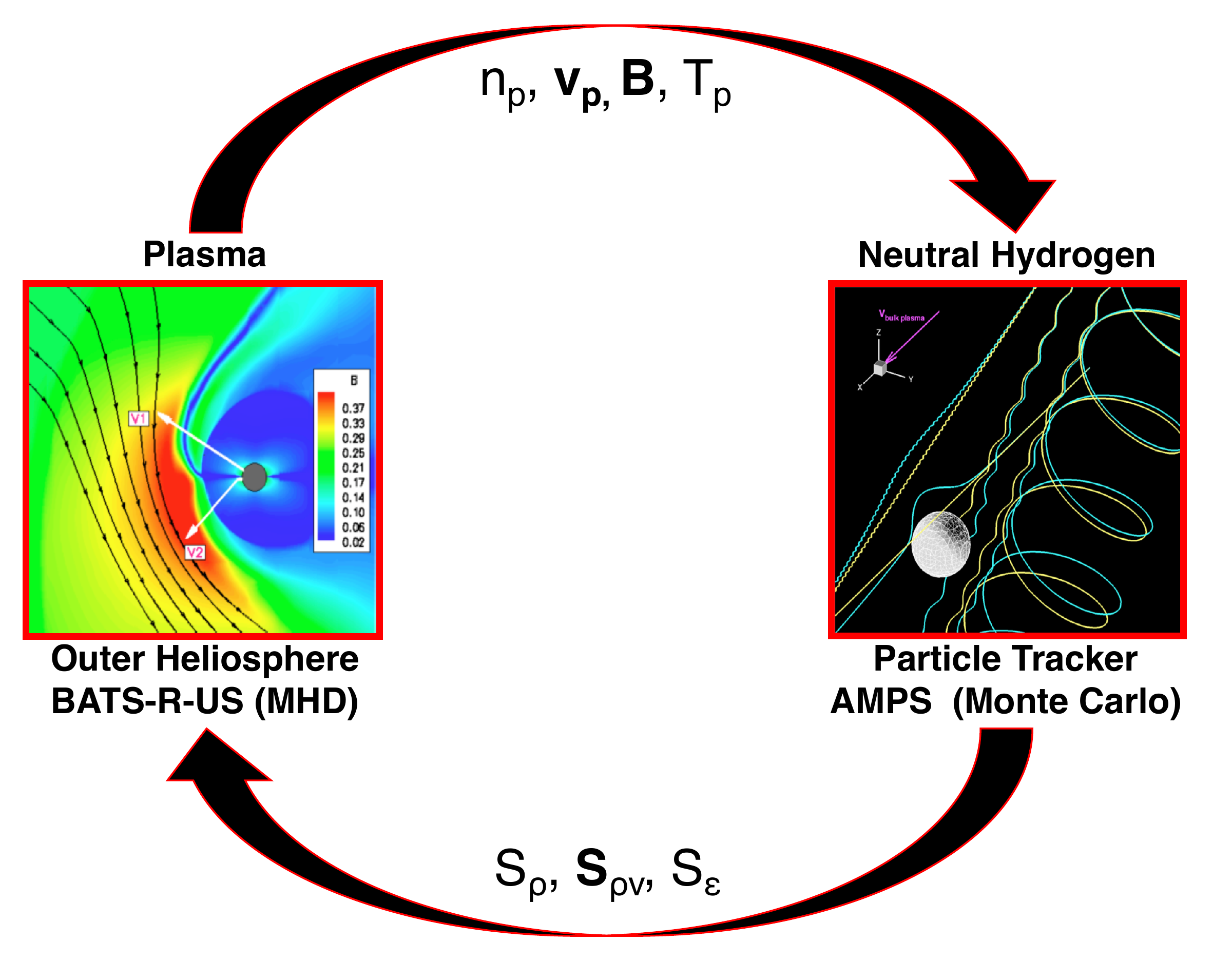}
\end{center}
\vspace{-10pt}
\caption{A schematic diagram summarizing the coupling between the OH and PT components within the SWMF. n, $\mathbf{v}_p$, \textbf{B}, and T$_p$ are the density, velocity, magnetic field, and temperature of the plasma. S$_\rho$, \textbf{S}$_{\rho v}$, S$_\epsilon$ are the source terms to the continuity, momentum and energy equations. \label{fig:ohptcoupling}}
\end{figure}

Due to the efficiency of analytic approximation for the source terms, a steady-state solution from the multi-fluid model is used as input to the SHIELD model. This brings the plasma closer to the final solution and reduces the amount of time needed to run the kinetic-MHD model. Figure \ref{fig:ohptcoupling} details the coupling between the Monte-Carlo and MHD models over the course of one time step in the plasma fluid. The SHIELD model passes the MHD variables from BATS-R-US to AMPS. AMPS then pushes the neutral particles through the plasma solution. The neutrals are injected with a Maxwell-Boltzmann distribution from the east face of the domain. AMPS determines when and where charge exchange will occur for each particle, given the local plasma conditions. When an event occurs, an ENA is generated and the resulting change in momentum and energy of the neutral is added to the local plasma source terms within that cell. The source terms are then passed back to BATS-R-US and added to the momentum and energy equations. The plasma solution is then advanced a time step and the updated solution is then passed back to AMPS. This process continues until the solution relaxes into a new steady state.

In the SHIELD model, the source terms must be updated every time-step of the plasma. The MHD solution cannot be advanced with constant source terms from a previous solution for an extended period of time as was implemented by \citet{baranov1993} and also used by \citet{alouani2011}. If the plasma solution is advanced in time with constant source terms from a previous solution then the SHIELD model develops a numerical instability in the flanks of the heliosphere. This was also seen by \citet{heerikhuisen2006}. Waves form at the outer boundary, pulling the HS plasma out into the ISM. This instability does not occur in the \citet{baranov1993} model due to their treatment of the HP as a perfectly ideal, tangential discontinuity that does not allow the HS plasma to leak into the ISM.

The source terms are calculated over a time interval, $\Delta t$. For dynamic problems, the coupling frequency, and therefore $\Delta t$, is set by the minimum temporal scale resolved in the MHD solution. This ensures the source terms are updated frequently enough to resolve all desired phenomena. For short timescales, this requires a large number of simulated macro-particles to obtain the necessary statistics for smooth source terms. Steady-state problems are time-independent such that the terms in Equation \ref{eq:Boltzmann} and the MHD equations do not vary with time. These runs can be handled differently than dynamic problems since we do not need to worry about resolving any temporal phenomena. The SHIELD model, can therefore advance the Monte-Carlo and MHD models at different rates, allowing AMPS to run for a longer period of time to develop smooth source terms before they are updated in the MHD equations and the plasma solution advanced. A similar method was implemented by \citet{heerikhuisen2006}. This reduces the need for a very large amount of macro-particles. The SWMF controls the frequency with which each component advances a time step. The SHIELD model cycles the components such that the MHD model can be restricted to advance a single time step over the $\Delta t$ interval of the kinetic code. 

Additional techniques can also be applied to time-independent problems to increase efficiency and reduce numerical costs. Charge exchange in the heliosphere produces neutral populations with very different characteristic speeds. Setting a single time step for all particles can have challenges. In order to control the quality of the statistical sample, the time-step is chosen such that particles cannot move through multiple cells within a single step. However, a time-step that is too short could cause a large number of particles to build up when slower particles reside in the largest cells of non-uniform grids. This can cause the computational nodes to run out of memory. Ideally, particles should spend three iterations within a cell for the model to build up sufficient statistics. Therefore, we have allowed steady-state simulations to be able to separate the neutral hydrogen atoms into four species, determined according to their energy. Each neutral hydrogen species is allowed to have its own time step. This ensures that the aforementioned problems do not occur and allows the solution to relax to a steady state much sooner, significantly reducing the execution time of the model.

\subsection{Model Comparison}
There are no direct observations of the neutral hydrogen atoms in the outer heliosphere to validate the SHIELD model. Neither \textit{Voyager 1} nor \textit{Voyager 2} took direct measurements of the neutral atoms. ENA maps also present a challenge for code validation. ENAs measured by IBEX and Cassini are produced primarily through neutral charge exchange with PUIs \citep{zank2010} and are line of sight measurements. While dynamic solar wind boundary conditions are included within our model \citep{michael2015}, our MHD model does not include the PUIs as a separate population. This makes a direct comparison to ENA observations an ineffective way to verify the correct inclusion of the kinetic model.   

In order to ensure the SHIELD model is coupled correctly, we compare our simulation to the results of the outer heliosphere kinetic-MHD models of \citet{alexashov2005} and \citet{heerikhuisen2006}. Both of theses models are 2D, axisymmetric, hydrodynamic simulations that treat the thermal solar wind ions, PUIs, and electrons as a single fluid. Each neglect photoionization and electron impact ionization, therefore, the neutrals interact with the plasma only through charge exchange, as was done with the SHIELD model. In this work, we adopt the same boundary conditions for the solar wind and ISM as \citet{alexashov2005} and \citet{heerikhuisen2006}. We also neglect the magnetic field, as done in the prior works. Our inner boundary is located at 30 au for the plasma unlike the other models which institute the boundary at 1 au. The solar wind conditions at 30 au were extracted from \citet{alexashov2005} using the WebPlotDigitizer Software\footnote{https://apps.automeris.io/wpd/} and taken to be constant and uniform in both latitude and longitude with $n_{SW}= 0.008$ cm$^{-3}$, $V_{SW}= 354.75$ km s$^{-1}$, $T_{SW}= 1.126\times10^5$ K. The ISM plasma enters the domain with the conditions: n$_{p_{ISM}}=0.06$ cm$^{-3}$, T$_{ISM}=6527$ K, and V$_{ISM}=26.4$ km s$^{-1}$ parallel to the ecliptic plane. The interstellar neutrals are injected with a Maxwell-Boltzmann distribution with a density of n$_{H_{ISM}}=0.18$ cm$^{-3}$ and the same bulk velocity and temperature of the plasma. 

\citet{alexashov2005} and \citet{heerikhuisen2006} both used the charge exchange cross section from \citet{mahertinsley1977} given by
\begin{equation}
    \sigma_{ex}(v_{rel}) = (1.6-0.0695 ln(v_{rel}))^2 10^{-14} cm^2
    \label{eq:MTsigma}
\end{equation}
This cross section is similar to Equation \ref{eq:LSsigma} from \citet{lindsay2005} for energies below $\sim10$ keV. For consistency, we have used the \citet{mahertinsley1977} cross section for this comparison. The relative velocity, $v_{rel}$, is given by 
\begin{equation}
    v_{rel} = v_{th,p} \left (\frac{e^{-\omega^2}}{\sqrt{\pi}} + \left ( \omega+\frac{1}{2\omega}\right )erf(\omega) \right ), ~\omega=\frac{|\mathbf{v}_H-\mathbf{u}|}{v_{th,p}}
    \label{eq:vrel}
\end{equation}
as done in \citet{heerikhuisen2006}.

SHIELD is a 3D model. Its computational domain extends from -1000 au to 1000 au in all 3 directions. The ISM enters into the domain from the $x=-1000$ au face. In the nose of the heliosphere we use a grid resolution of 4 au inside a -400 au to 400 au cube around the Sun to resolve the hydrogen wall and the disturbed ISM. This grid was used for both the MHD and kinetic models to ensure consistency between the multi-fluid and kinetic approach. The mean free path of the neutral atoms is on the order of 100 au \citep{izmodenov2015}, therefore a 4 au resolution is ample to resolve the neutral hydrogen solution within AMPS. Since our model is 3-dimensional instead of 2-dimensional, we will qualitatively compare the solution in the meridional cut of our 3D simulation to the works of \citet{alexashov2005} and \citet{heerikhuisen2006}. However, a 2D asymmetric model with the correct boundary conditions should yield the same solution as its 3D counterpart. This comparison is done for both the multi-fluid model, using the four neutral fluid approximation, as well as the SHIELD model.

The multi-fluid approximation is more computationally efficient than the K-MHD model, therefore it is used to relax the plasma to a steady solution. \citet{alexashov2005} showed that the four neutral fluid approximation, most closely matched the solution run with their K-MHD model. In an effort to obtain the plasma solution closest to the true solution to reduce the execution time of the SHIELD model, we also describe the neutrals with four separate neutral fluids. The neutral populations are separated as follows. Pristine ISM neutrals or population IV neutrals, enter the domain with values set by the boundary conditions. For these conditions, a bow shock forms within the solution. Population IV neutrals are defined as being supersonic with flow speed less than 140 km s$^{-1}$, distinguishing them from particles born in the supersonic solar wind. The region between the BS and the HP is the disturbed ISM, or Population I neutrals, defined to have a sonic Mach number less than one and a temperature below $7.5\times10^4$ K. Population II neutrals in the heliosheath occur where the plasma is subsonic at temperatures higher than Population I neutral hydrogen. Finally, population III neutrals are generated in the supersonic solar wind. A steady solution for the multi-fluid model is reached after the model is advanced for over 50,000 time steps. 

\section{Results}
As stated in the introduction, a consensus has not been reached amongst the most recent 3D kinetic-MHD models \citep{izmodenov2015,pogorelov2013} on the exact impact the solar magnetic field has on the global structure of the heliosphere. Additionally, they employ different numerical algorithms at the heliopause. The varying plasma solutions between the models offers little in verifying that we have coupled the kinetic and MHD models correctly. We, therefore, rely on the comparison to the simplified, hydrodynamic simulations of \citet{alexashov2005} for model verification, as done in \citet{heerikhuisen2006}. Future studies will explore a more detailed comparison of the 3D kinetic-MHD models.

\subsection{Fixed Plasma Solution}
\begin{figure}[!ht]
\centering
\subfloat[]{%
  \includegraphics[width=0.32\textwidth]{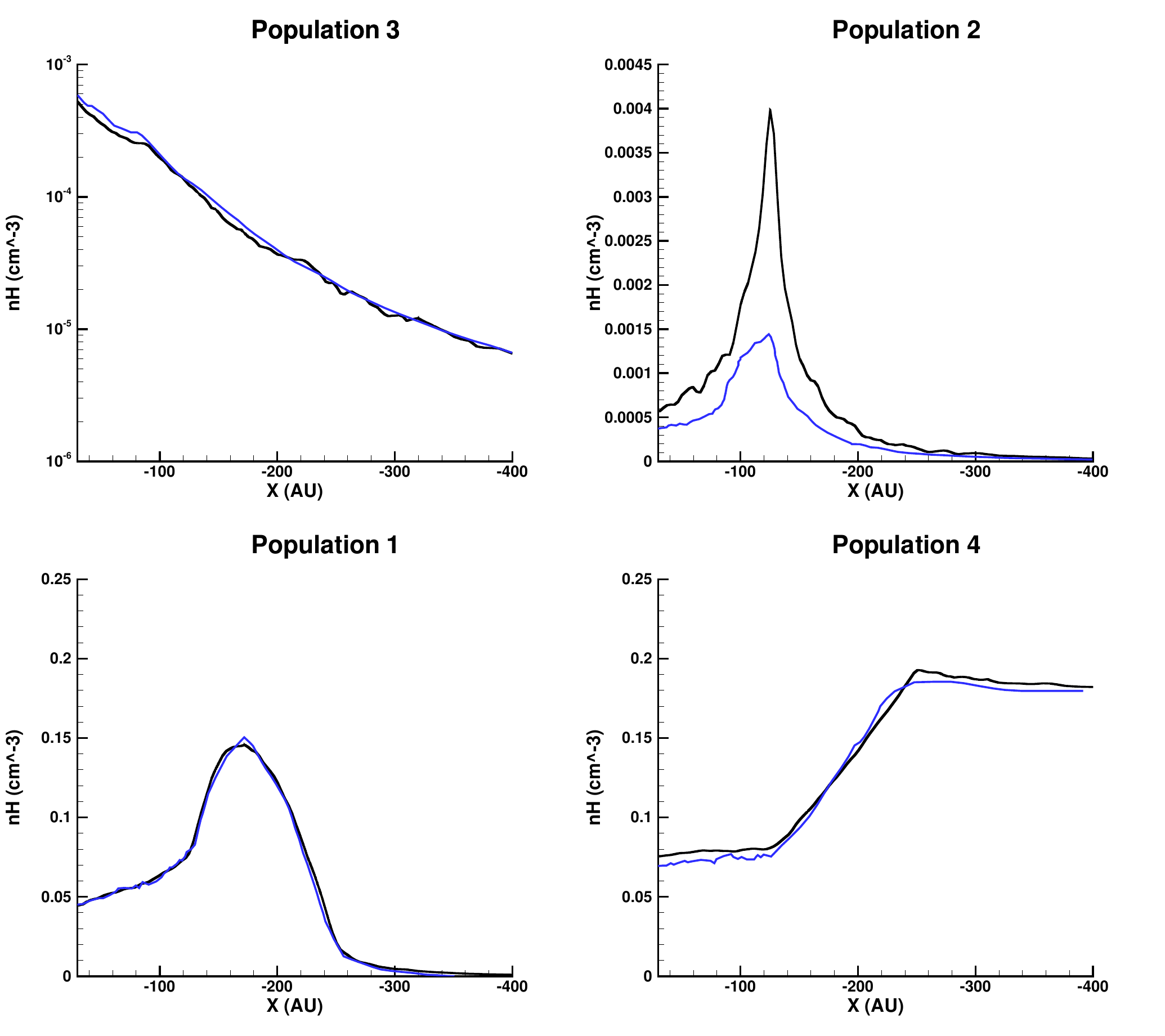}
  \label{fig:FixedPlasma_nH}%
}
\subfloat[]{%
  \includegraphics[width=0.32\textwidth]{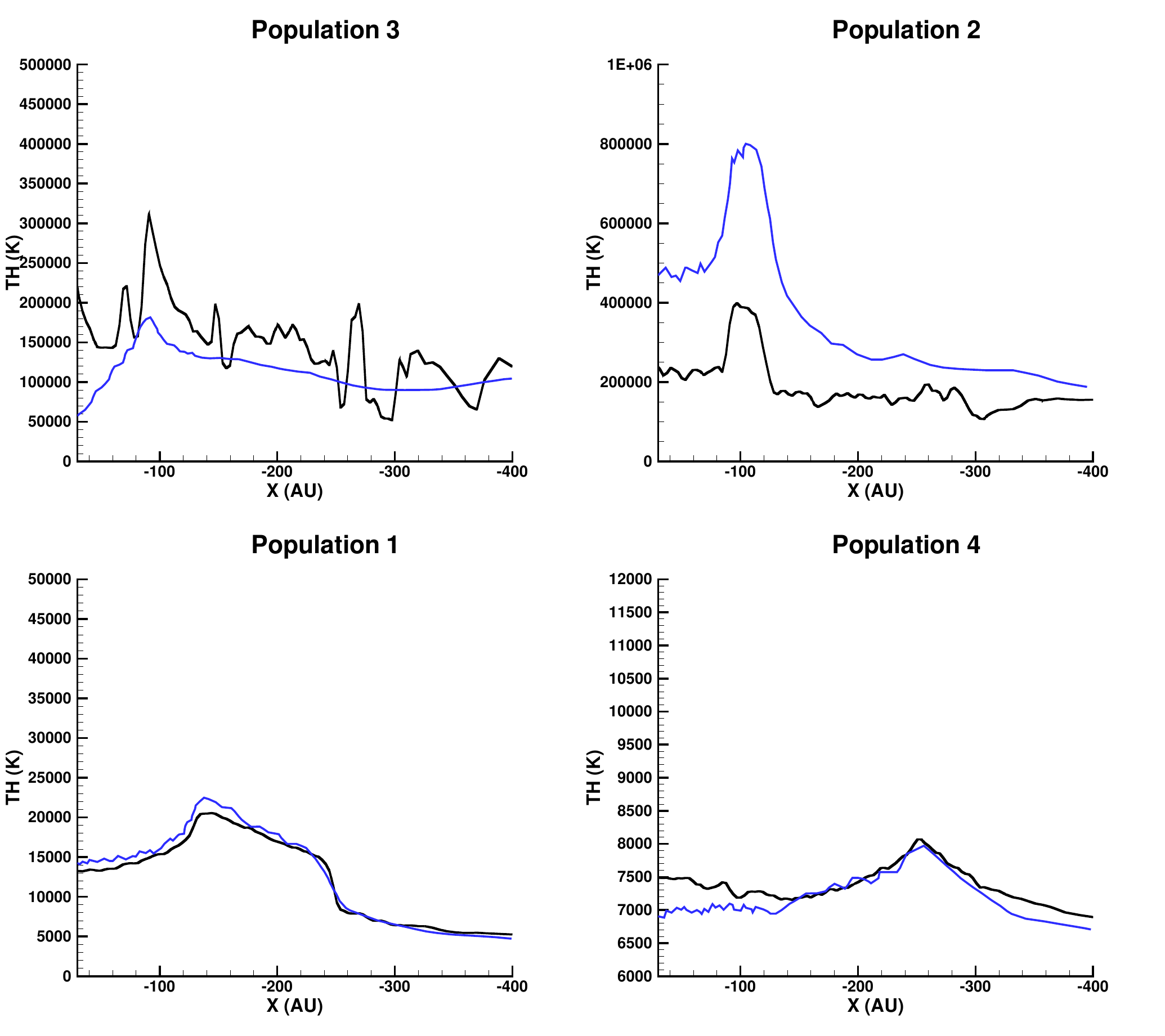}
  \label{fig:FixedPlasma_TH}%
}
\subfloat[]{%
  \includegraphics[width=0.32\textwidth]{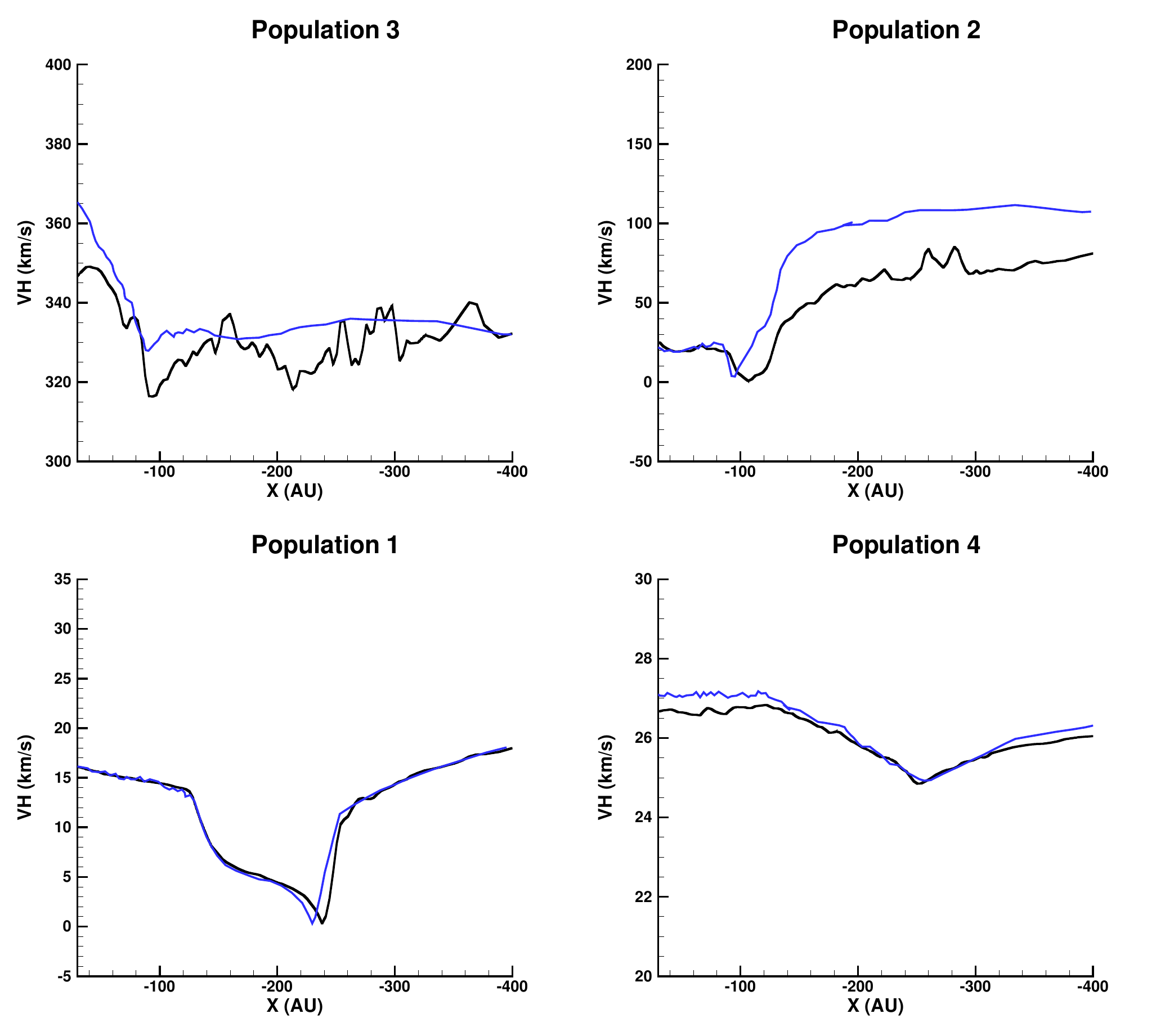}
  \label{fig:FixedPlasma_VH}%
}
\caption{The neutral density [cm$^{-3}$] \protect\subref{fig:FixedPlasma_nH}, temperature [K] \protect\subref{fig:FixedPlasma_TH}, and speed [km/s] \protect\subref{fig:FixedPlasma_VH} of the four neutral populations along the nose of the heliosphere for a fixed plasma solution. The blue curve shows the results of \citet{alexashov2005} and the black curve are of the SHIELD model.
\label{fig:FixedPlasma}}
\end{figure}

It is important to ensure that the kinetic solution predicts the correct neutral solution before analyzing the fully coupled self-consistent SHIELD model. We, therefore, first look at how the neutral solution differs from the work of \citet{alexashov2005} using a fixed plasma solution. For a fixed plasma solution, the MHD model and Monte Carlo code are one-way coupled. The plasma solution is passed to the kinetic code allowing the kinetic solver to generate a neutral solution but does not allow the source terms from charge exchange to be added to the MHD equations and alter the plasma. AMPS is therefore able to simulate transport of the neutral particles through the plasma, allowing for the generation of ENAs, while holding the plasma solution constant. The plasma solution used was taken from the steady state of the fully coupled SHIELD model and the neutral populations within the SHIELD model were selected with the same criteria as the multi-fluid model described previously. Figure \ref{fig:FixedPlasma} compares the neutral density, speed, and temperature for the kinetic solution of the SHIELD model to the kinetic solution of \citet{alexashov2005} for each population of neutrals. The numbers refer to the respective region of the heliosphere where they were created. In our scheme Populations 4, 1, 2, and 3 correspond to neutrals born in the pristine ISM, between the BS and HP, heliosheath, and the supersonic solar wind, respectively. 

A comparison of our kinetic solution to that of \citet{alexashov2005} allows us to verify that our kinetic solver is determining the correct distribution for the interstellar neutrals. The data from \citet{alexashov2005} was extracted using the WebPlotDigitizer Software\footnote{https://apps.automeris.io/wpd/}.  \citet{alexashov2005}  use a different numbering scheme for their neutral populations, referring the pristine ISM, the region between the BS and HP, the heliosheath, and the supersonic solar wind as populations 4,3,2, and 1, respectively. Their populations have been renamed to match the definitions used in this work. \citet{alexashov2005} used the fixed plasma solution from their 2D self-consistent K-MHD model. Although similar boundary conditions were used, the hydrogen wall in our 3D multi-fluid steady state model is almost 100 au thicker than the fixed plasma solution used by \citet{alexashov2005}. Despite this difference, the kinetic solutions shows very good qualitative and quantitative agreement. 

Both kinetic models predict higher filtration of interstellar neutral populations, 4 and 1, into the heliosphere than their multi-fluid counter parts. The higher filtration of neutral particles into the heliosphere occurs within the kinetic model due to the large mean free path of hydrogen atoms. There are no collisions between hydrogen atoms therefore the large charge exchange mean free path ($> 100-200$ au) allows kinetic neutral hydrogen atoms to travel into the upwind direction from the sides or flanks of the heliosphere. The collisions between neutrals, dominates in the fluid description, causing the neutrals in the multi-fluid model to not propagate as far for particles that would be crossing flow streamline. This causes the multi-fluid model to more closely follow the streamlines. Less contribution from neutrals in the flanks of the heliosphere causes the multi-fluid model to predict a lower neutral density in the nose of the heliosphere. 

Populations 1, 3, and 4 match very well between the kinetic models throughout the entirety of the domain in both value and behavior. AMPS predicts a denser, cooler neutral population originating in the HS than \citet{alexashov2005}. This population, however, is very sensitive to how the HP is defined when separating the populations. The difference in our solutions is most likely due to a different definition separating hydrogen wall neutrals from neutrals born in the HS. If some of the neutrals originating in the hydrogen wall are included in HS population, the population will appear to be denser and cooler as in the SHIELD model. Overall the kinetic solution within SHIELD matches the results of \citet{alexashov2005} very well and we can be confident that AMPS is providing an accurate representation of the interstellar neutrals as the propagate through the heliosphere.

Additionally, the kinetic treatment shows that the average speed of neutrals originating in the the pristine ISM (population 4) increases as they propagate into the heliosphere. Faster interstellar neutrals have larger charge exchange mean-free-paths as they travel through the hydrogen wall. As a result, slower particles undergo charge exchange more frequently. Charge exchange removes the slower neutrals from population 4 causing an increase in the average speed of the population. 

\subsection{Fully Coupled Self-Consistent Kinetic Solution}
The multi-fluid solution is used to start the fully coupled, self-consistent SHIELD model. The plasma solution from the multi-fluid model is passed to the SHIELD model. Neutrals are then propagated through the domain to generate a self-consistent kinetic solution that can be used to update the plasma with the corresponding source terms. The problem is then treated as time independent and the neutral hydrogen atoms are modeled as four separate ENA species. The OH is cycled such that the MHD solution is advanced one time step for every interval of time that the source terms are calculated over within the PT component. This allows statistics to accumulate over a long period of time in between each individual step in the plasma solution to reduce the number of model particles needed. We model just under 100 million particles and allow the statistics for the source terms to accrue for 10,000 time steps before the resulting source terms are passed back to the MHD solver and the plasma solution is updated and passed back to AMPS. The cycling procedure continues until the plasma relaxes to a new solution This occurred after the MHD solution was advanced 700 time steps. 

\begin{figure}[htpb]
\centering
\subfloat{%
  \includegraphics[width=0.32\textwidth]{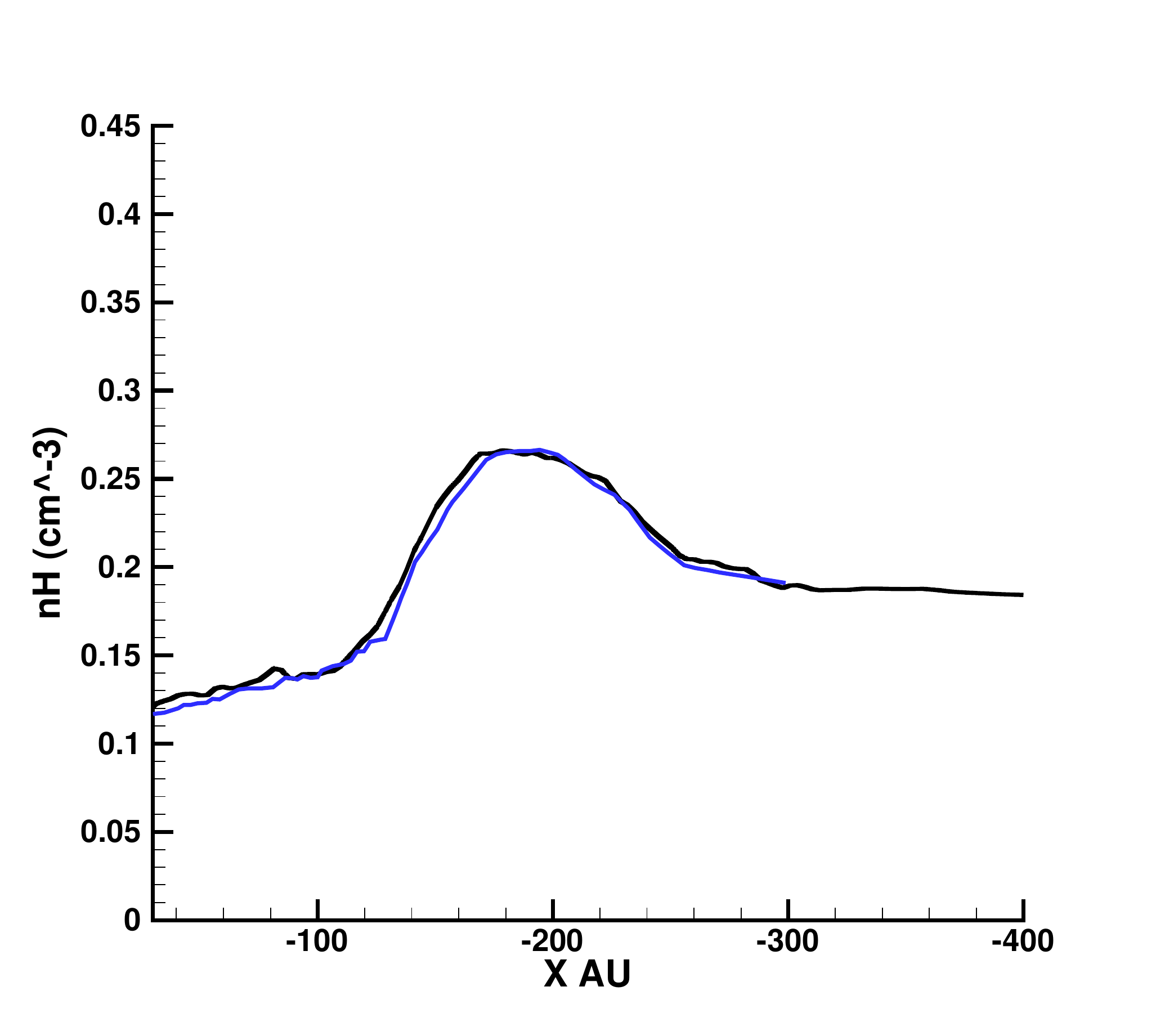}
}
\subfloat{%
  \includegraphics[width=0.32\textwidth]{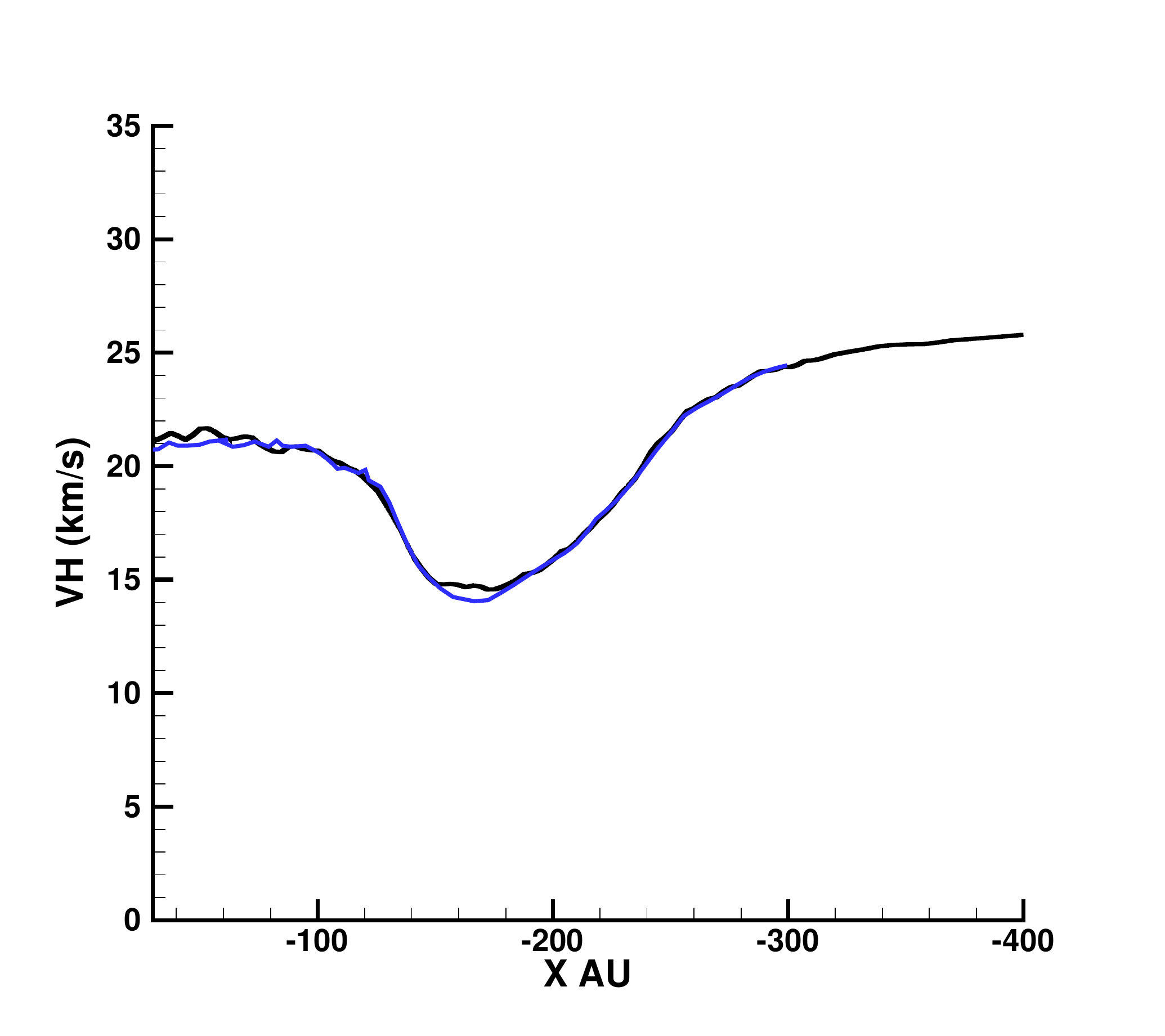}
}
\subfloat{%
  \includegraphics[width=0.32\textwidth]{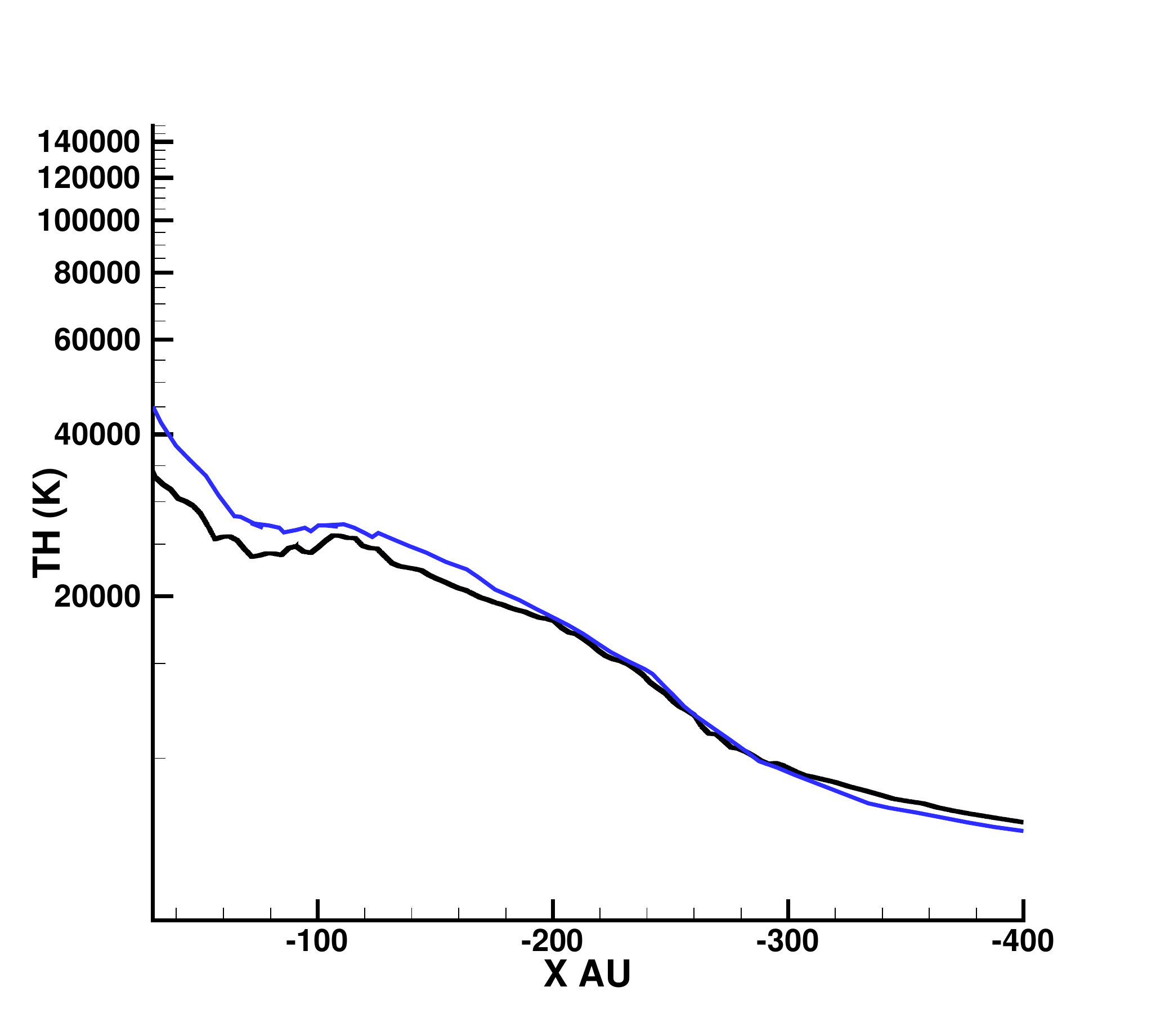}
}
\caption{The neutral density, speed and temperature in the upwind direction for the fully coupled, self-consistent kinetic models. The results of SHIELD are presented black while the results of \citet{alexashov2005} are shown in blue.}  
\label{fullyCoupled_neutral}
\end{figure}

\begin{figure}[htpb]
\centering
\subfloat{%
  \includegraphics[width=0.32\textwidth]{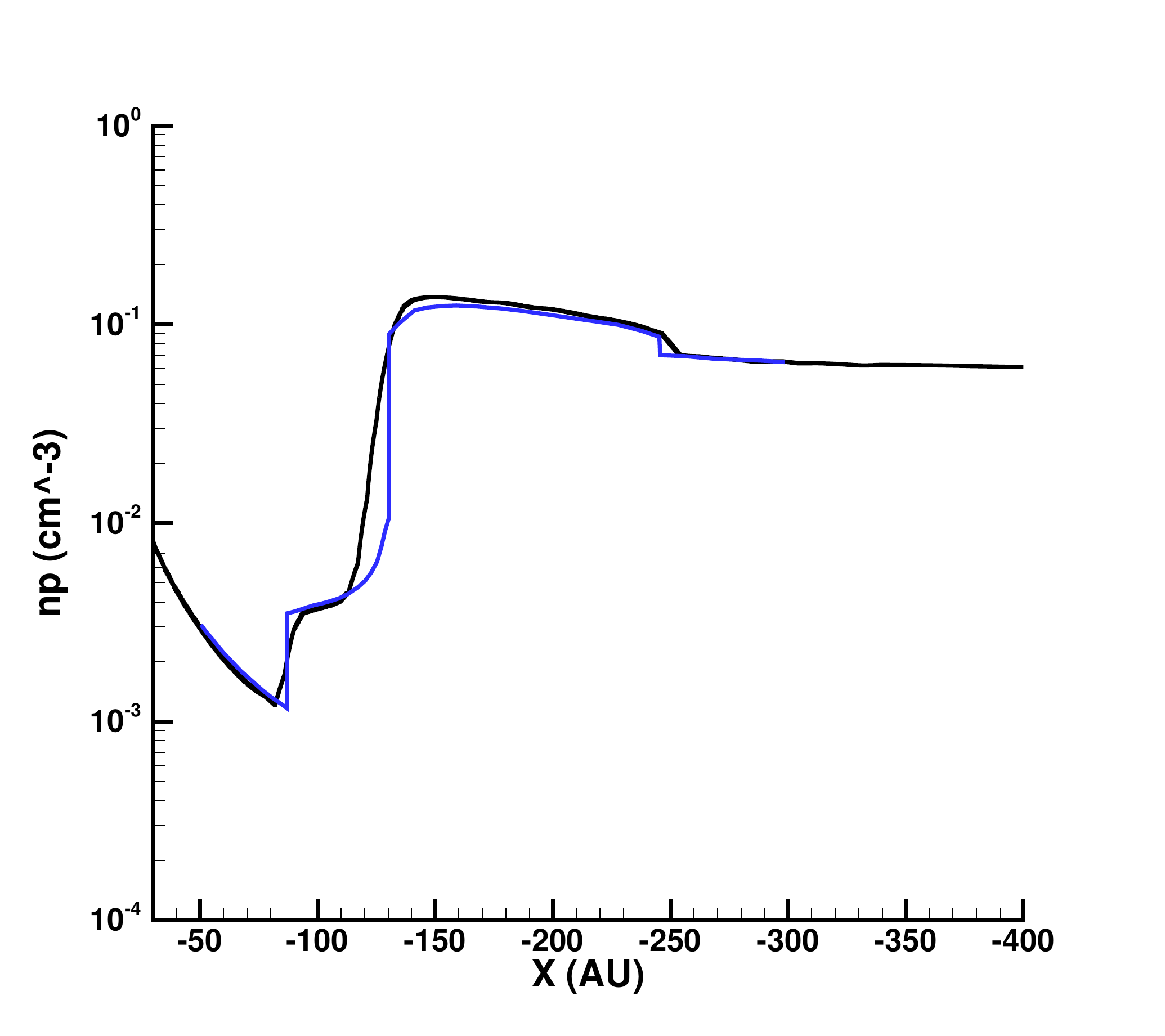}
}
\subfloat{%
  \includegraphics[width=0.32\textwidth]{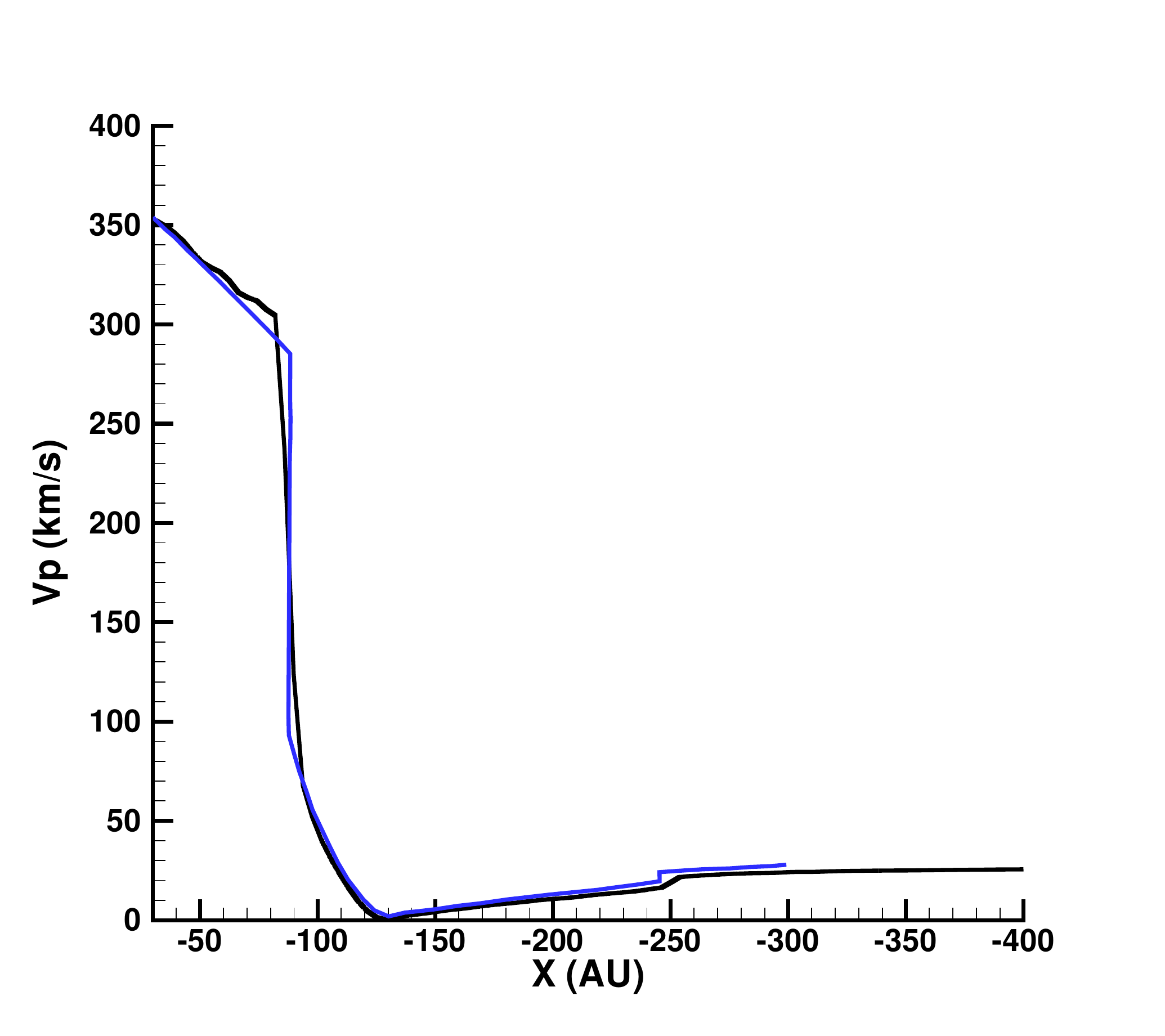}
}
\subfloat{%
  \includegraphics[width=0.32\textwidth]{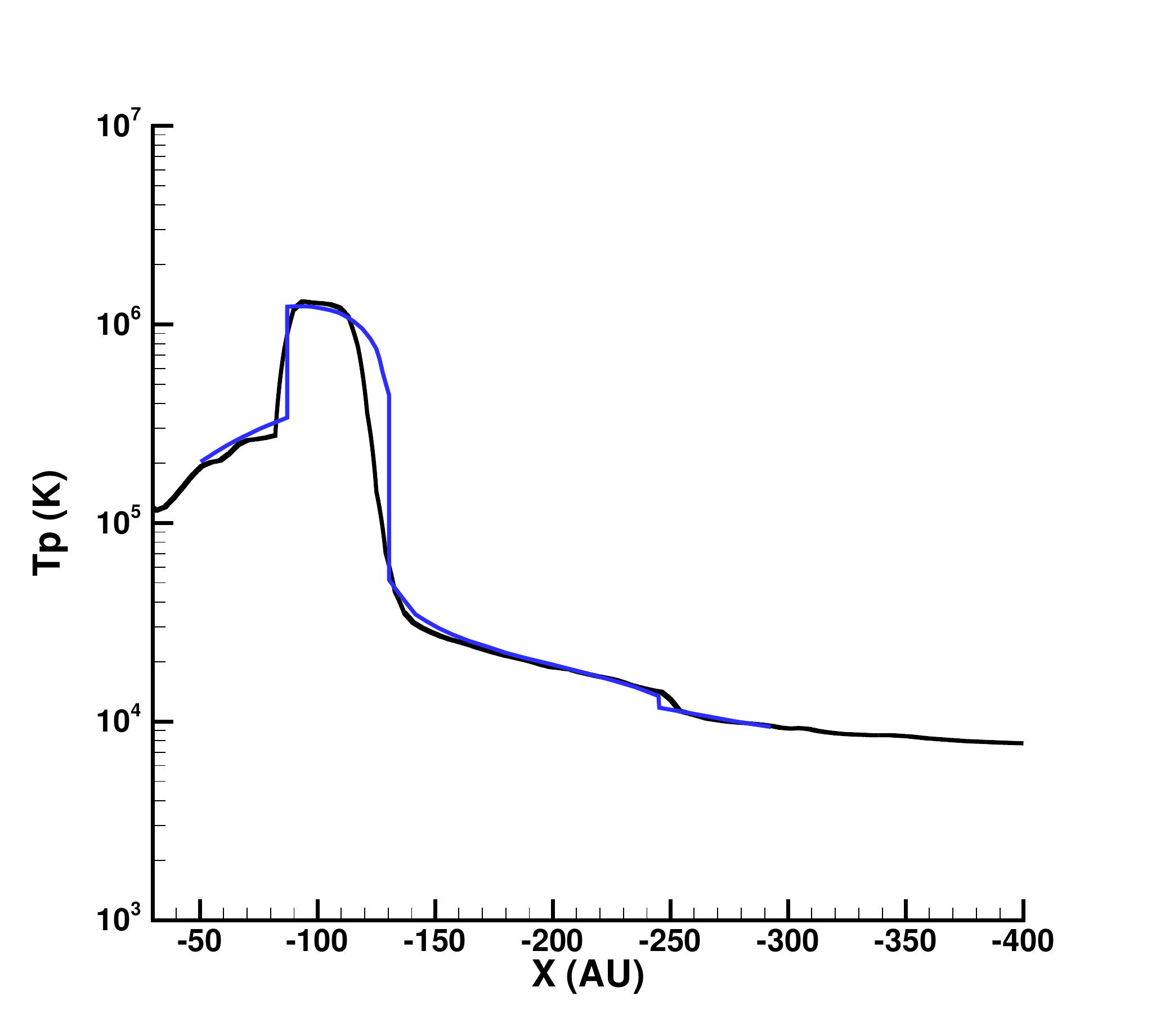}
}
\caption{The plasma density, speed, and temperature in the upwind direction for the fully coupled, self-consistent kinetic model. The results of SHIELD are in black while the results of \citet{alexashov2005} are shown in blue.}  
\label{fullyCoupled_plasma}
\end{figure}

Figure \ref{fullyCoupled_neutral} and Figure \ref{fullyCoupled_plasma} compare 1D profiles in the nose of the heliosphere from the fully coupled, self-consistent results from SHIELD for both the neutrals and the plasma, respectively. The kinetic solution of \citet{alexashov2005} are also shown. In comparing SHIELD to the well-established model results of \citet{alexashov2005}, we see overall excellent agreement between the two approaches. Both models qualitatively predict similar structures within the neutral solution. The kinetic models have very good quantitative agreement at the peak of the hydrogen wall in the density, speed, and temperature of the neutrals as well as the values within the heliosphere. Both kinetic solutions predict denser, slower population of neutrals propagating through the heliosphere in comparison to their multi-fluid counterparts. The kinetic treatment of the neutrals is a significant improvement of our model to match the neutral solution of \citet{alexashov2005}.

Furthermore, SHIELD predicts the same penetration of neutrals into the heliosphere, which will act to increase charge exchange within the supersonic solar wind and HS. The main feature seen in both K-MHD models is that the filtration of neutrals into the heliosphere is 20\% higher than their respective multi-fluid neutral models. A comparison of the plasma solutions predicted by the models is presented in Figure \ref{fullyCoupled_plasma}. The SHIELD model predicts the TS to move 5 au towards the Sun as a result of the additional charge exchange from the higher density of interstellar neutrals within the heliosphere in comparison to the multi-fluid model. 

\begin{figure}[htpb]
\centering
\subfloat[]{%
  \includegraphics[width=0.5\textwidth]{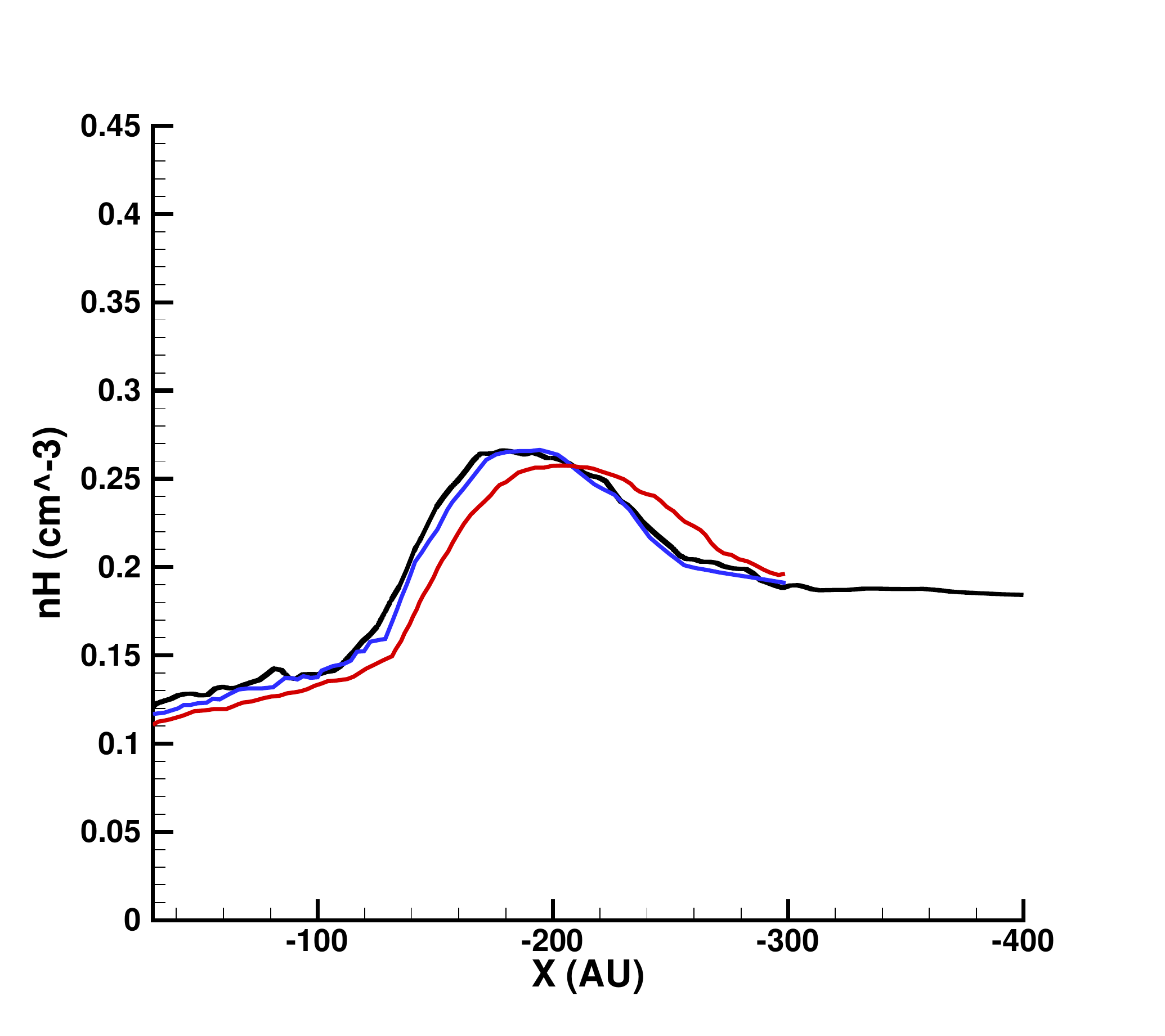}
  \label{fig:nH_SHIELD_HZ06_AI05}%
}

\subfloat[]{%
  \includegraphics[width=0.5\textwidth]{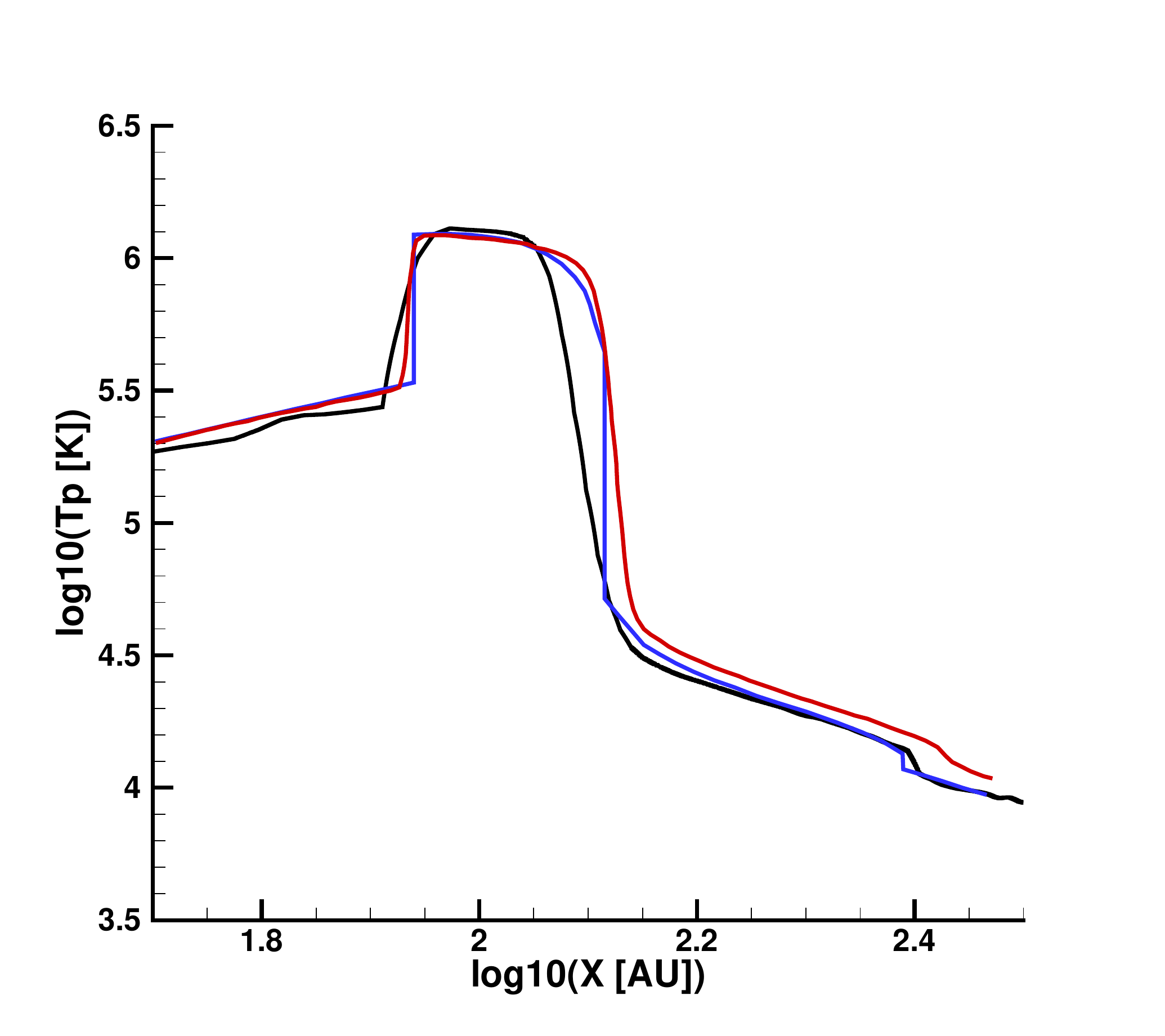}
  \label{fig:Tp_SHIELD_HZ06_AI05}%
} 
\caption{Neutral density \protect\subref{fig:nH_SHIELD_HZ06_AI05} and plasma temperature \protect\subref{fig:Tp_SHIELD_HZ06_AI05} profile along the nose of the heliosphere. The plots compare the results of the kinetic models of \citet{alexashov2005} (blue) and \citet{heerikhuisen2006} (red) to the SHIELD model (black).}  
\label{fig:SHIELD_HZ06_AI05}
\end{figure}

\citet{heerikhuisen2006}, performed a similar run when developing the K-MHD model within the Multi-scale Fluid-Kinetic Simulation Suite. The TS, HP and BS location within all three kinetic simulations can be seen Figure \ref{fig:Tp_SHIELD_HZ06_AI05} and are presented in Table \ref{tab:ModelBoundaryLocations}. Here, the HP location in the SHIELD model is determined by velocity streamlines. Both \citet{alexashov2005} and \citet{heerikhuisen2006} observe the kinetic neutrals to cause the TS to move 5-6 au towards the Sun due to the increased hydrogen filtration into the heliosphere. The SHIELD model predicts this as well although has a final TS 2-3 au further than the other kinetic models. It should be noted that our multi-fluid model also saw a similar discrepancy in the TS location as well. Overall the SHIELD model matches the heliospheric locations of \citet{alexashov2005} very well. 

Figure \ref{fig:SHIELD_HZ06_AI05} directly compares the model results from the SHIELD model to those of \citet{alexashov2005} and \citet{heerikhuisen2006} along the nose of the heliosphere. The SHIELD model shows excellent agreement with both models. The SHIELD model closely reproduces the neutral density profile in \citet{alexashov2005}. The BS in the SHIELD model is 5 au further than in \citet{alexashov2005}, leading to a slightly larger hydrogen wall thickness. All three models predict similar neutral density filtration into the heliosphere. A comparison of the plasma temperature in the kinetic models is presented in Figure \ref{fig:Tp_SHIELD_HZ06_AI05}. The TS, HP, and BS locations can be seen within the models. The SHIELD model predicts a slightly cooler supersonic solar wind but shows very good agreement in the HS and in the shocked ISM, predicting a BS in between the other two models. One of the main differences between the SHIELD model and the model of \citet{alexashov2005} is the discontinuity algorithm used in their model. This can be seen by the vertical lines in their solution at the HP. The solution within SHIELD shows a much more gradual decrease from the temperature within the HS to the disturbed ISM ahead of the HP. This gradient is also seen in the model of \citet{heerikhuisen2006}, however, it appears more evident in the SHIELD model due to a slightly coarser resolution in the area. The largest discrepancy within the models occurs in between the BS and the HP. The SHIELD model predicts similar neutral hydrogen density and plasma temperature to \citet{alexashov2005}. The model of \citet{heerikhuisen2006} predicts a hotter region that pushes the BS further from the Sun. This results in a wider hydrogen wall whose density peaks 30 au further away from the HP than the SHIELD model. Overall, the SHIELD model has excellent agreement with the results of \citet{alexashov2005}, validating that the PT component was applied correctly to the outer heliosphere and produces a correct solution that we can trust moving forward.

\section{Summary}
In this work, we presented the SHIELD model. A 3D, global, self-consistent K-MHD model of the outer heliosphere implemented within SWMF. The SHIELD model couples AMPS, a Monte Carlo code, to our MHD solver to treat the neutrals streaming through the heliosphere kinetically. SHIELD models the thermal protons, PUIs, and electrons as a single plasma fluid and and includes hydrogen as the only neutral species. Here, H-H collisions, photoionization, and radiation pressure are neglected. Therefore, the only way the plasma and neutrals interact is through charge exchange. 

We compare the results of SHIELD to the results of \citet{alexashov2005} to verify that the kinetic treatment of the neutrals was incorporated correctly. Overall, we find excellent agreement in the neutral distributions between the models. SHIELD can accurately reproduce the behavior of the each population of neutrals as they propagate through the system and matches the higher filtration of the neutrals into the heliosphere. The SHIELD model accurately reproduces the locations of the heliospheric boundaries predicted by the K-MHD results of \citet{alexashov2005}, including the the TS moving inward by 6 au as well as the HP distance in the nose getting 10 au closer to the Sun. The SHIELD model is a significant improvement over our multi-fluid model in matching the results of other models. The excellent agreement between the models make us confident that SHIELD is coupled correctly and can be used for future work investigating the heliosphere. 

The plasma is also strongly affected by the different modeling techniques of the neutrals. The plasma solution is the SHIELD model shows excellent agreement with the results of \citet{alexashov2005}. As with other works, we conclude that the differences between the kinetic distributions and the multi-fluid approach are large and that the kinetic solution should be used when comparing model results to observations. 

\acknowledgments
This work was supported by NASA Headquarters under the NASA Earth and Space Science Fellowship Program - Grant NNX14AO14H. A. M. and M. O. acknowledge the support of NASA Grand Challenge NNX14AIB0G and the the Hariri fellowship from Boston University. The calculations were performed at NASA AMES Pleiades Supercomputer. 


\begin{thebibliography}{}
\bibitem[Alexashov, \& Izmodenov(2005)]{alexashov2005} Alexashov, D., \& Izmodenov, V.\ 2005, \aap, 439, 1171
\bibitem[Alexashov et al.(2004)]{alexashov2004} Alexashov, D.~B., Chalov, S.~V., Myasnikov, A.~V., et al.\ 2004, \aap, 420, 729
\bibitem[Alexashov et al.(2016)]{alexashov2016} Alexashov, D.~B., Katushkina, O.~A., Izmodenov, V.~V., et al.\ 2016, \mnras, 458, 2553
\bibitem[Alouani-Bibi et al.(2011)]{alouani2011} Alouani-Bibi, F., Opher, M., Alexashov, D., Izmodenov, V., \& Toth, G.\ 2011, \apj, 734, 45
\bibitem[Baranov et al.(1981)]{baranov1981} Baranov, V.~B., Ermakov, M.~K., \& Lebedev, M.~G.\ 1981, Soviet Astronomy Letters, 7, 206
\bibitem[Baranov, \& Malama(1993)]{baranov1993} Baranov, V.~B., \& Malama, Y.~G.\ 1993, \jgr, 98, 15157
\bibitem[Bird(1994)]{bird1994} Bird, G.~A.\ 1994, Molecular Gas Dynamics And The Direct Simulation Of Gas Flows
\bibitem[Borovikov et al.(2017)]{borovikov2017} Borovikov, D., Tenishev, V., Gombosi, T.~I., et al.\ 2017, \apj, 835, 48
\bibitem[Dong et al.(2018)]{dong2018} Dong, C., Bougher, S.~W., Ma, Y., et al.\ 2018, Journal of Geophysical Research (Space Physics), 123, 6639
\bibitem[Drake et al.(2015)]{drake2015} Drake, J.~F., Swisdak, M., \& Opher, M.\ 2015, \apjl, 808, L44
\bibitem[Fahr et al.(2000)]{fahr2000} Fahr, H.~J., Kausch, T., \& Scherer, H.\ 2000, \aap, 357, 268
\bibitem[Glocer et al.(2009)]{glocer2009} Glocer, A., T{\'o}th, G., Ma, Y., et al.\ 2009, Journal of Geophysical Research (Space Physics), 114, A12203
\bibitem[Gloeckler \& Geiss(2001)]{gloeckler2001} Gloeckler, G., \& Geiss, J.\ 2001, Joint SOHO/ACE Workshop ``solar and Galactic Composition'', 281
\bibitem[Gruntman(2015)]{gruntman2015} Gruntman, M.\ 2015, Journal of Geophysical Research (Space Physics), 120, 6119
\bibitem[Heerikhuisen et al.(2006)]{heerikhuisen2006} Heerikhuisen, J., Florinski, V., \& Zank, G.~P.\ 2006, Journal of Geophysical Research (Space Physics), 111, A06110
\bibitem[Izmodenov et al.(2004)]{izmodenov2004} Izmodenov, V., Malama, Y., Gloeckler, G., et al.\ 2004, \aap, 414, L29
\bibitem[Izmodenov et al.(2003)]{izmodenov2003} Izmodenov, V., Malama, Y.~G., Gloeckler, G., et al.\ 2003, \apjl, 594, L59
\bibitem[Izmodenov, \& Alexashov(2015)]{izmodenov2015} Izmodenov, V.~V., \& Alexashov, D.~B.\ 2015, \apjs, 220, 32
\bibitem[Izmodenov et al.(2001)]{izmodenov2001} Izmodenov, V.~V., Gruntman, M., \& Malama, Y.~G.\ 2001, \jgr, 106, 10681
\bibitem[Izmodenov et al.(2000)]{izmodenov2000} Izmodenov, V.~V., Malama, Y.~G., Kalinin, A.~P., et al.\ 2000, \apss, 274, 71
\bibitem[Krimigis et al.(2009)]{krimigis2009} Krimigis, S.~M., Mitchell, D.~G., Roelof, E.~C., et al.\ 2009, Science, 326, 971
\bibitem[Lee et al.(2014)]{lee2014} Lee, Y., Combi, M.~R., Tenishev, V., et al.\ 2014, Journal of Geophysical Research (Planets), 119, 905
\bibitem[Lindsay, \& Stebbings(2005)]{lindsay2005} Lindsay, B.~G., \& Stebbings, R.~F.\ 2005, Journal of Geophysical Research (Space Physics), 110, A12213
\bibitem[Maher, \& Tinsley(1977)]{mahertinsley1977} Maher, L.~J., \& Tinsley, B.~A.\ 1977, \jgr, 82, 689
\bibitem[Malama(1991)]{malama1991} Malama, Y.~G.\ 1991, \apss, 176, 21
\bibitem[McNutt et al.(1998)]{mcnutt1998} McNutt, R.~L., Lyon, J., \& Goodrich, C.~C.\ 1998, \jgr, 103, 1905
\bibitem[Michael et al.(2015)]{michael2015} Michael, A.~T., Opher, M., Provornikova, E., et al.\ 2015, \apjl, 803, L6
\bibitem[Michael et al.(2018)]{michael2018} Michael, A.~T., Opher, M., \& T{\'o}th, G.\ 2018, \apj, 860, 171
\bibitem[M{\"o}bius et al.(2012)]{mobius2012} M{\"o}bius, E., Bochsler, P., Bzowski, M., et al.\ 2012, \apjs, 198, 11
\bibitem[M{\"u}ller et al.(2008)]{muller2008} M{\"u}ller, H.-R., Florinski, V., Heerikhuisen, J., et al.\ 2008, \aap, 491, 43
\bibitem[M{\"u}ller et al.(2000)]{muller2000} M{\"u}ller, H.-R., Zank, G.~P., \& Lipatov, A.~S.\ 2000, \jgr, 105, 27419
\bibitem[Opher et al.(2009)]{opher2009} Opher, M., Bibi, F.~A., Toth, G., et al.\ 2009, \nat, 462, 1036
\bibitem[Opher et al.(2011)]{opher2011} Opher, M., Drake, J.~F., Swisdak, M., et al.\ 2011, \apj, 734, 71
\bibitem[Opher et al.(2017)]{opher2017} Opher, M., Drake, J.~F., Swisdak, M., et al.\ 2017, \apjl, 839, L12
\bibitem[Opher et al.(2015)]{opher2015} Opher, M., Drake, J.~F., Zieger, B., et al.\ 2015, \apjl, 800, L28
\bibitem[Opher et al.(2003)]{opher2003} Opher, M., Liewer, P.~C., Gombosi, T.~I., et al.\ 2003, \apjl, 591, L61
\bibitem[Opher et al.(2004)]{opher2004} Opher, M., Liewer, P.~C., Velli, M., et al.\ 2004, \apj, 611, 575
\bibitem[Opher et al.(2019)]{opher2019} Opher, M., Loeb, A., Drake, J., et al.\ 2019, Nature Astronomy, Accepted
\bibitem[Pauls et al.(1995)]{pauls1995} Pauls, H.~L., Zank, G.~P., \& Williams, L.~L.\ 1995, \jgr, 100, 21595
\bibitem[Pogorelov et al.(2009)]{pogorelov2009} Pogorelov, N.~V., Borovikov, S.~N., Florinski, V., et al.\ 2009, Numerical Modeling of Space Plasma Flows: ASTRONUM-2008, 149
\bibitem[Pogorelov et al.(2013)]{pogorelov2013} Pogorelov, N.~V., Suess, S.~T., Borovikov, S.~N., et al.\ 2013, \apj, 772, 2
\bibitem[Pogorelov \& Zank(2005)]{pogorelov2005} Pogorelov, N.~V., \& Zank, G.~P.\ 2005, Advances in Space Research, 35, 2055
\bibitem[Provornikova et al.(2013)]{provornikova2013} Provornikova, E., Opher, M., Izmodenov, V., et al.\ 2013, \aap, 552, A99
\bibitem[Provornikova et al.(2014)]{provornikova2014} Provornikova, E., Opher, M., Izmodenov, V.~V., et al.\ 2014, \apj, 794, 29
\bibitem[Richardson(2008)]{richardson2008} Richardson, J.~D.\ 2008, \grl, 35, L23104
\bibitem[Ruci{\'n}ski et al.(1996)]{rucinski1996} Ruci{\'n}ski, D., Cummings, A.~C., Gloeckler, G., et al.\ 1996, \ssr, 78, 73
\bibitem[Schwadron et al.(2011)]{schwadron2011} Schwadron, N.~A., Allegrini, F., Bzowski, M., et al.\ 2011, \apj, 731, 56
\bibitem[Schwadron et al.(2014)]{schwadron2014} Schwadron, N.~A., Moebius, E., Fuselier, S.~A., et al.\ 2014, \apjs, 215, 13
\bibitem[Tenishev et al.(2008)]{tenishev2008} Tenishev, V., Combi, M., \& Davidsson, B.\ 2008, \apj, 685, 659
\bibitem[Tenishev et al.(2014)]{tenishev2014} Tenishev, V., Rubin, M., Shou, Y., et al.\ 2014, Lunar and Planetary Science Conference, 1305
\bibitem[T{\'o}th et al.(2005)]{toth2005} T{\'o}th, G., Sokolov, I.~V., Gombosi, T.~I., et al.\ 2005, Journal of Geophysical Research (Space Physics), 110, A12226
\bibitem[T{\'o}th et al.(2012)]{toth2012} T{\'o}th, G., van der 
Holst, B., Sokolov, I.~V., et al.\ 2012, Journal of Computational Physics, 
231, 870
\bibitem[Valeille et al.(2009)]{valeille2009} Valeille, A., Combi, M.~R., Bougher, S.~W., et al.\ 2009, Journal of Geophysical Research (Planets), 114, E11006
\bibitem[Wallis(1975)]{wallis1975} Wallis, M.~K.\ 1975, \nat, 254, 202
\bibitem[Williams et al.(1997)]{williams1997} Williams, L.~L., Hall, D.~T., Pauls, H.~L., et al.\ 1997, \apj, 476, 366
\bibitem[Zank et al.(2010)]{zank2010} Zank, G.~P., Heerikhuisen, J., Pogorelov, N.~V., et al.\ 2010, \apj, 708, 1092
\bibitem[Zank et al.(1996)]{zank1996} Zank, G.~P., Pauls, H.~L., Williams, L.~L., et al.\ 1996, \jgr, 101, 21639
\end{thebibliography}

\begin{table}
\caption{Distances (in au) to the TS, HP, and BS in the nose of the heliosphere}
\centering
\begin{tabular}{l c c c c c}
\hline
   \textbf{Model} & \textbf{Termination Shock} & \textbf{Heliopause} & \textbf{Bow Shock} \\ 
\hline
   SHIELD  & 89 & 128 & 250   \\
   A\&I & 87 & 130 & 245 \\
   HFZ & 86 & 140 & 262 \\
\hline
\end{tabular}
\\
A\&I denotes the work of \citet{alexashov2005} and HFZ that of \citet{heerikhuisen2006}. Values are shown for the self-consistent, kinetic-MHD models from the respective works.
\label{tab:ModelBoundaryLocations}
\end{table}

\end{document}